\documentclass[times,astrobib,amssymb]{mn2e}
\usepackage{psfig}
\newcommand{\lapp}{\mbox{\raisebox{-0.3em}{$\stackrel{\textstyle <}{\sim}$}}}
\newcommand{\gapp}{\mbox{\raisebox{-0.3em}{$\stackrel{\textstyle >}{\sim}$}}}
\newcommand{\be}{\begin{equation}}
\newcommand{\en}{\end{equation}}

\def\zabs{$z_{\rm abs}$}
\def\zem{$z_{\rm em}$~}

\def\mgii{Mg~{\sc ii} }
\def\feiia{Fe~{\sc ii}$\lambda$2600}
\def\mgia{Mg~{\sc i}$\lambda$2852}
\def\mgiia{Mg~{\sc ii}$\lambda$2796}
\def\mgiib{Mg~{\sc ii}$\lambda$2803}
\def\mgiiab{Mg~{\sc ii}$\lambda\lambda$2796,2803}

\def\kms{km~s$^{-1}$}
\title[Intermediate redshift 21-cm absorbers]
{A complete sample of 21-cm absorbers at $z\sim 1.3$: Giant Metrewave Radio Telescope Survey Using Mg II Systems}
\author[Gupta et al.]{N. Gupta$^{1}$\thanks{E-mail: Neeraj.Gupta@atnf.csiro.au}, R. Srianand $^2$, P. Petitjean $^3$, P. Noterdaeme $^2$, D. J. Saikia $^4$\\
$^{1}$ Australia Telescope National Facility, CSIRO, Epping, NSW 1710, Australia \\
$^{2}$ IUCAA, Ganeshkhind, Pune 411007, India \\
$^{3}$ Universit\'e Paris 6, UMR 7095, Institut d'Astrophysique de Paris-CNRS, 98bis Boulevard Arago, 75014 Paris, France \\
$^{4}$ NCRA-TIFR, Ganeshkhind, Pune 411007, India }

\begin{document}

\date{Accepted. Received; in original form }

\pagerange{\pageref{firstpage}--\pageref{lastpage}} \pubyear{2009}

\maketitle

\label{firstpage}

\begin{abstract}
We present the results of a systematic Giant Metrewave Radio Telescope 
(GMRT) survey of 21-cm absorption in a representative and 
unbiased sample of 35 strong Mg~{\sc ii} systems in the redshift range: 
\zabs$\sim$1.10$-$1.45, 33 of which have $W_{\rm r}\ge$1\,\AA.  
The survey using $\sim$400\,hrs of telescope time has resulted in 
9 new 21-cm detections and stringent 21-cm optical depth upper limits 
(median 3$\sigma$ optical depth per 10\,km\,s$^{-1}$ of 0.017)
for the remaining 26 systems.
This is by far the largest number of 21-cm detections from any single survey of intervening absorbers. 
Prior to our survey no intervening 21-cm system was known in the above redshift
range and only one system was known in the redshift range
$0.7\le z \le 1.5$. 
We discuss the relation between 
the detectability of 21-cm absorption and various 
properties of UV absorption lines. We show that if Mg~{\sc ii} systems
are selected with the following criteria, Mg~{\sc ii} doublet ratio $\le$1.3 and  
$W_{\rm r}$(Mg~{\sc i})/$W_{\rm r}$(Mg~{\sc ii})~$\ge$~0.3, then 
a detection rate of 21-cm absorption up to 90\% can be achieved.  
We estimate $n_{21}$, the number per unit redshift of 21-cm absorbers with 
$W_r$(Mg~{\sc ii})~$>$~$W_{\rm o}$ and integrated optical depth 
$\cal{T}_{\rm 21}$ ~$>$~$\cal{T}_{\rm o}$ 
and show that $n_{21}$ decreases with increasing redshift. In particular, for $W_{\rm o}$~=~1.0~\AA~ 
and $\cal{T}_{\rm o} >$ ~0.3~km~s$^{-1}$,
$n_{21}$ falls by a factor 4 from $<z>$~=~0.5 to $<z>$~=~1.3. 
The evolution seems to be stronger for stronger Mg~{\sc ii} systems.  
Using a subsample of systems for which high frequency VLBA images are available, we show that 
the effect is not related to the structure of the background radio sources and 
is most probably due to the evolution of the cold neutral medium filling factor in Mg~{\sc ii} systems.
We find no correlation between the velocity spread of the 21-cm absorption feature
and $W_{\rm r}$(Mg~{\sc ii}) at $z\sim 1.3$. 
%
%
\end{abstract}
%

\section{Introduction}
Observations of high-$z$ galaxies suggest that 
the global comoving star-formation rate density 
peaks at $1\le z\le 2$ and then sharply decreases
towards $z\sim0$ (e.g. Madau et al. 1996, Hopkins 2004). 
The determination of the mass density of the gas and its content (molecules, dust and cold H~{\sc i} gas) 
over the same redshift range provides an independent and complementary understanding of the redshift 
evolution of star-formation at similar epochs.
While the H~{\sc i} content of galaxies can be best probed
by surveys of 21-cm emission, limited sensitivity 
of current radio telescopes does not allow them to reach beyond the 
local Universe (Zwaan et al. 2005 for $z\sim$0; 
Verheijen et al. 2007 and Catinella et al. 2008 for direct 
detections at $z\sim$0.2; Lah et al. 2007 for
statistical detection of H~{\sc i} at $z\sim0.24$). On the contrary, detection
of H~{\sc i} in the spectra of distant QSOs in the form of 
damped Lyman-$\alpha$ absorption provides a luminosity unbiased way 
of probing the evolution of the H~{\sc i} content in the universe 
(Lanzetta et al. 1991; Wolfe et al. 1995; Storrie-Lombardi et al. 1996; 
Ellison et al. 2001; Peroux et al. 2003; Prochaska, Herbert-Ford \& Wolfe, 
2005; Rao, Turnshek \& Nester, 2006; Prochaska \& Wolfe 2008; 
Noterdaeme et al. 2009).
%
%

A major improvement in the DLA statistics at $z\gapp2$
has been achieved by the Sloan Digital Sky survey (SDSS).
Spectroscopic catalogs are available for DR5 
(see Prochaska \& Wolfe 2008) and for DR7 (see Noterdaeme et al. 2009).
For lower redshifts,
initial detections of DLAs at $z<2$ were performed 
using the International Ultraviolet Explorer (IUE) and the Hubble Space Telescope (HST,
Lanzetta, Wolfe \& Turnshek, 1995; Jannuzi et al. 1998).
However, a large number of low-$z$ DLAs 
have been identified
after candidate DLAs are pre-selected through the presence of
strong metal absorption lines (Rao, Turnshek \& Nestor, 2006; RTN06
from now on).

It is believed that physical conditions 
in the neutral gas associated with normal galaxies are influenced by the 
radiative and mechanical feedback from the in-situ star formation activity (McKee \& Ostriker 1977).
Thus, it is a good idea to investigate whether there is 
any relationship between the evolution of the star-formation rate density
and the physical conditions in the H~{\sc i} gas probed by DLAs.
Our understanding of physical conditions in DLAs at $z\gapp2$ is 
largely based on the analysis of H$_2$ and/or atomic fine-structure transitions.  
A systematic search for molecular hydrogen in DLAs at high redshifts 
($z_{\rm abs}\gapp1.8$), using Ultraviolet and Visual Echelle Spectrograph
(UVES) at the Very Large Telescope (VLT) down to a detection limit of 
typically $N($H$_2)\sim 2\times 10^{14}$ cm$^{-2}$, has resulted in
a detection in $\sim$10-15\% of the cases (Ledoux, Petitjean \& Srianand 2003; 
Noterdaeme et al 2008a). HD and CO molecules are detected in, respectively,
two and one DLAs 
(Varshalovich et al. 2001; Noterdaeme et al. 2008b, Srianand et al. 2008a). 
The rotational excitation of the molecules and fine-structure excitations 
of C~{\sc i} and C~{\sc ii} in these systems are consistent with the absorption lines 
originating from dense cold gas in radiative equilibrium with the star light 
produced by the in-situ star-formation (Srianand et al. 2000; Ge, Bechtold \& Kulkarni 
2001; Reimers et al. 2003; Srianand et al. 2005; Hirashita \& Ferrara 2005; Cui et al. 2005; 
Noterdaeme et al. 2007a, 2007b). Absence of H$_2$ in most of the
high-$z$ DLAs is consistent with the gas being either warm neutral or warm ionized 
(Petitjean, Srianand \& Ledoux 2000; Srianand et al. 2005). 
The absorption lines of C~{\sc ii$^*$} are detected in all the H$_2$ bearing DLAs
(Srianand et al. 2005) and in roughly 50\% of the DLAs without H$_2$ 
(Wolfe, Prochaska \& Gawiser 2003; Wolfe, Gawiser \& Prochaska 2003;
Srianand et al. 2005; Wolfe et al. 2008).  As C~{\sc ii$^*$} is an important coolant,
its column density combined with $N$(H~{\sc i}) can be used to 
discuss the cooling rate in the absorbing gas and therefore the SFR
associated with the DLA. Unfortunately for the time being
the above mentioned tracers can not be used to probe the
physical state of the absorbing gas at $z\lapp1.8$ because the 
useful transitions are located below the atmospheric cut-off.

Most of our initial understanding of the physical state of H~{\sc i} gas in the 
Galactic interstellar medium (ISM) is based on the 21-cm absorption line
(see Kulkarni \& Heiles 1988 for details). 
It is widely believed that the spin temperature ($T_{\rm s}$) of the H~{\sc i} 
gas is a reliable tracer of the kinetic temperature (see Roy, Chengalur \& Srianand 2006 
for direct confirmation). 
The H~{\sc i} column density for an optically thin cloud that covers a 
fraction $f_{\rm c}$ of the background radio source is related to the 21-cm optical 
depth $\tau(v)$ in a velocity interval $v$$-$$v+$d$v$ and to the 
spin temperature ($T_{\rm s}$) by (e.g. Kulkarni \& Heiles 1988)
\begin{equation}
N{\rm(H~I)}=1.835\times10^{18}~{T_{\rm s}\over f_{\rm c}}\int~\tau(v)~{\rm d}v~{\rm cm^{-2}}.
\label{eq1}
\end{equation}
In addition, the width of the 21-cm line detected in a high resolution 
spectrum yields a direct measurement of (or stringent upper limit on) 
the kinetic temperature.
Thus detecting 21-cm absorption is important to probe the physical 
conditions in the interstellar medium of galaxies at intermediate redshifts.   

RTN06 have shown that DLAs essentially 
have Mg~{\sc ii} rest equivalent width, $W_{\rm r}$(\mgiia)$\ge$0.6\AA.  
Therefore, the search of 21-cm absorption in a sample of strong Mg~{\sc ii} absorbers 
is an unique way 
to probe the redshift evolution of physical conditions in DLAs 
like absorption systems 
at low-$z$.  
In the next Section we summarize the previous efforts to search for 21-cm absorption in DLAs 
and \mgii systems. Due to various practical reasons 
(e.g. frequency coverage of receivers, radio frequency interference, RFI)
the redshift coverage is sparse and measurements are available only 
for a few systems, especially in the redshift range: $1\le z\le 2$.
The Giant Meterwave Radio Telescope (GMRT), with its very sensitive 610
MHz receiver, is the only radio telescope available at present in the {\it relatively} RFI-clean 
environment ({\bf e.g.} compared to Green Bank Telescope or Westerbork Synthesis Radio Telescope)
for covering part of this redshift range.
In addition, the large number of publicly available high-$z$ QSO spectra from SDSS
allow one to construct the large samples of Mg~{\sc ii} absorbers suitable for the 
searches of DLAs/21-cm absorbers (cf. Section~3).
Motivated by this we have conducted a 
systematic search for 21-cm absorption in a representative sample of Mg~{\sc ii}
systems in the redshift range $1.10 \le z \le 1.45$ and discovered 9 new 21-cm absorbers. 
This is by far the highest number of detections from any single survey of intervening 
21-cm absorption (see Srianand, Gupta \& Petitjean 2007; Gupta et al. 2007 and 
Srianand et al. 2008b for preliminary results from the survey).
%

This paper is structured as follows.
We give a brief summary of the past 21-cm searches performed by other groups 
in Section~2. In Section~3, we describe our sample and details of 
observations are presented in Section~4. 
In Section~5, we describe in detail the individual 21-cm absorbers 
detected in our survey. Detectability of 21-cm absorbers and its 
relationship to metal absorption equivalent widths, equivalent width
ratios and structure of QSOs at radio wavelengths are discussed in 
Section~6.  In Section~7, we derive the number 
density per unit redshift of 21-cm absorbers.
Using the low-$z$ sample of Lane (2000) we investigate the redshift
evolution of the number density of 21-cm absorbers.
In Section~8, we explore the relationship between the velocity spread of 
21-cm absorption and  the equivalent width of the \mgii absorption line.
Discussion of our results along with a summary are presented in 
Section~9.

\begin{figure}
\centerline{\hbox{
\psfig{figure=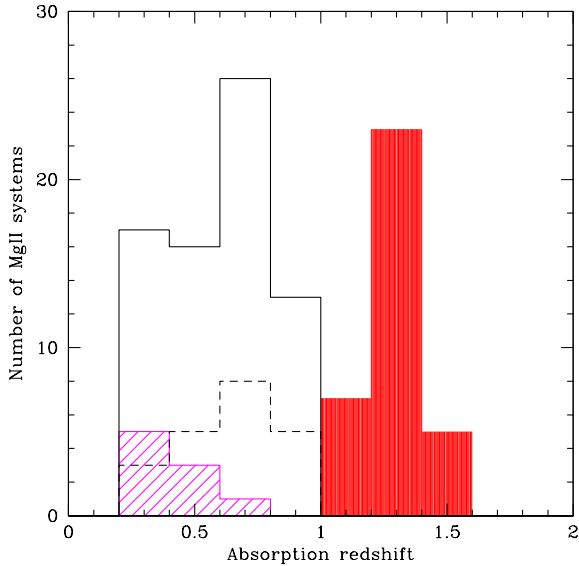,height=8.0cm,width=8.0cm,angle=0}
}}
\caption[]{Redshift distribution of Mg~{\sc ii} systems that were
searched for 21-cm absorption. The filled histogram is the GMRT
sample of 35 Mg~{\sc ii} systems presented in this paper (33 of these absorption 
systems have $W_{\rm r}$(Mg~{\sc ii}$\lambda$2796)$\gapp1${\AA}). 
The solid line histogram is for the sample of Lane (2000). 
The hatched histogram corresponds to 21-cm detections in this sample.
The distribution for the $W_{\rm r}$(Mg~{\sc ii}$\lambda$2796)$\ge 1${\AA}
sub-set of these systems is given by the dashed line histogram. 
}
\label{zdist}
\end{figure}

\section{Brief summary of previous 21-cm absorption line surveys}

There have been several systematic surveys for 21-cm absorption in
DLAs and/or strong metal line absorption systems undertaken
by various groups over the past two decades.
Briggs \& Wolfe (1983) searched 18 Mg~{\sc ii} systems, with 
$W_{\rm r}$(Mg~{\sc ii})$\ge$0.5\AA~ and $0.3\le z\le 1.8$, and detected two.
Lane (2000), in her thesis work, searched 62 Mg~{\sc ii}
systems at $0.2\le z\le0.9$ without imposing any equivalent widths cut-off
and detected 21-cm absorption in 3 systems. 
Results of 21-cm searches are available for 7 DLAs at $z\ge2.8$ 
(Carilli et al. 1996; Kanekar \& Chengalur 2003) with only one 
detection (Kanekar et al. 2007). Measurements
are consistent with $T_{\rm s}$~$\ge$~1000\,K. There
are five 21-cm detections reported in the literature for
$1.7\le z\le2.8$ 
[\zabs = 2.289 towards TXS\,0311+430 (York et al. 2007);
\zabs = 2.347 towards B0438-436 (Kanekar et al. 2006);
\zabs = 2.039 towards B0458-020 (Wolfe et al. 1985); 
\zabs = 1.943 towards 1157+014 (Wolfe \&  Briggs, 1981); 
and
\zabs =1.776 towards 1331+170 (Wolfe \& Davis, 1979)].
But non-detections are not systematically reported in this redshift range.
At $z<0.9$, Kanekar \& Chengalur (2003) summarize 21-cm optical
depths for 14 known DLAs and 12 detections (including their own observations 
of two DLAs namely PKS\,1629+12 and PKS\,2128-13). 
Curran et al. (2007) report the detection of a
21-cm absorption at \zabs = 0.656 in the complex sight-line towards 3C336.
In total 18 intervening 21-cm absorption systems are known till now.

We plot in Fig.~\ref{zdist} the redshift distribution of the Mg~{\sc ii} absorption systems
surveyed by Lane (2000). This is  the only large survey at low-$z$ for which both detections 
and non-detections are systematically reported.
It includes 62 observations using WSRT and 10 other 
systems from the literature satisfying their selection criterion (see Lane 2000 for details).
The detections shown as a hatched histogram include
detections reported in Lane (2000) together with detections 
from better quality data by Kanekar \& Chengalur (2003) and 
Curran et al. (2007) for systems that were originally reported as 
non-detections. In the same figure, the filled histogram shows the distribution of 
Mg~{\sc ii} systems in our GMRT sample.  
For equivalent width cutoff of $\sim$ 1{\AA}, our GMRT sample
has more than twice the number of systems investigated by Lane (2000).
A detailed comparison of the results
from low and high redshift samples will be discussed in Section~7.
 
\section{Sample of \mgii systems in front of radio quasars}
\label{samp}

RTN06 have shown that DLAs can be preselected on the basis of  
the equivalent widths of Mg~{\sc ii}, Fe~{\sc ii} and Mg~{\sc i} 
absorption lines. Specifically, they found that 36\%$\pm$6\% of the 
Mg~{\sc ii} absorbers with Fe~{\sc ii}$\lambda$2600 and 
Mg~{\sc ii}$\lambda$2796 rest equivalent widths greater than 0.5 \AA~ are DLAs.  
%
\begin{figure}
\centerline{\hbox{
\psfig{figure=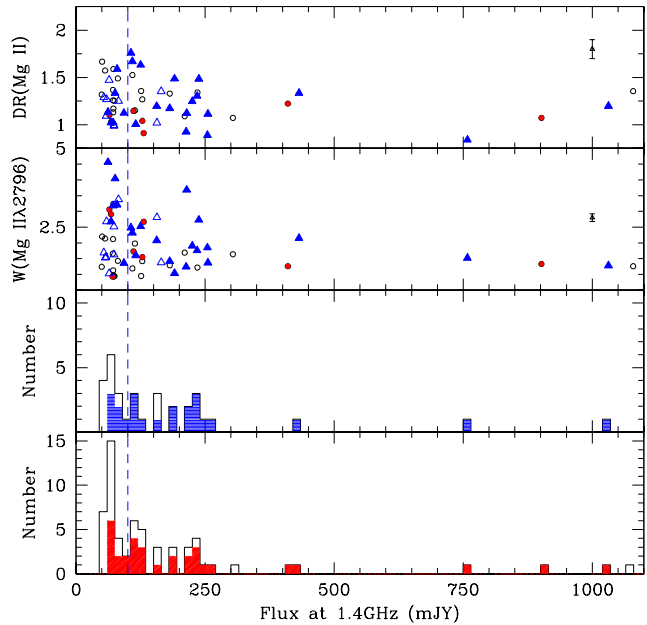,height=9.0cm,width=9.0cm,angle=0}
}}
\caption[]{
{\it Top panels:} The \mgii doublet ratios and Mg~{\sc ii}$\lambda$2796
equivalent widths in our sample are plotted
against the 1.4\,GHz peak flux density of the background source. 
Triangles are for systems from DR3 (P06) and circles are for 
additional systems 
in DR5. 
Filled and open symbols are for systems with and without GMRT 
observations. Typical error in the measurements are shown in these
panels.
{\it Third and bottom panel:} Distribution of 
radio flux densities in our sample. The shaded histograms gives 
the distributions of objects observed with GMRT.
}
\label{totsample}
\end{figure}
Motivated by this we have conducted a GMRT 21-cm absorption survey 
over the past two years in absorption systems 
selected on the basis of Mg~{\sc ii}$\lambda$2796 rest equivalent width
in the redshift range  1.10$\le z\le1.45$. The constraint on the
redshift range comes from the frequency range covered by the 
GMRT 610 MHz receiver. Our sample is drawn from 
the identification of Mg~{\sc ii} systems by Prochter, Prochaska \& Burles (2006, hereafter P06) 
in SDSS DR3 and by us using our automatic procedure for additional systems in DR5. We do not make any
further selection cut based on the equivalent widths of Fe~{\sc ii}  and
Mg~{\sc i} absorption lines. Thus our candidate selection is based on 
$W_{\rm r}$(Mg~{\sc ii}$\lambda$2796) alone.

There are 1953 \mgii systems in P06 with $W_{\rm r}$(Mg~{\sc ii}$\lambda$2796)
 $\ge 1.0$ {\AA} in the redshift range: 1.10$\le z \le 1.45$. 
By cross-correlating the corresponding QSO positions with the NRAO VLA Sky Survey (NVSS) radio catalog (Condon et al. 1998) 
we found 45 of these \mgii systems having background QSO with flux density at
1.4\,GHz in excess of 50\,mJy. We manually checked the SDSS spectra
of all these sources to confirm the \mgii systems and remeasure accurately the $\lambda$2796
rest equivalent width as well as other lines. 
This step is essential as the identifications of Mg~{\sc ii} systems
are not 100\% complete and errors in the Mg~{\sc ii} equivalent width
measurement can be large (see discussion on this in P06) when
one is using automatic procedures. We find that \mgii identification in one of
the QSOs (i.e \zabs = 1.109 towards the very bright radio source J104833.69+600845.7) 
is spurious.  
While manually checking the SDSS spectra we detected an additional 
(i.e. not catalogued in P06) Mg~{\sc ii} system at \zabs~=~1.3739 towards J123431.72+645556.5.  
Since this system has $W_{\rm r}$(Mg~{\sc ii}$\lambda$2796)~=~0.95$\pm$0.07, we included it in our sample 
as it satisfies the selection criteria of $W_{\rm r}\gapp1$\AA~ within 1~$\sigma$ 
and 1.10$\le$\zabs$\le$1.45. 

In order to minimize the possibility that the optical and radio sightlines 
are not identical we looked at the $\sim$5$^{\prime\prime}$ resolution radio images of the 44 remaining sources 
(with 45 absorption line systems) 
obtained in the Faint Images of the Radio Sky at Twenty-centimeters (FIRST) 
survey\footnote{FIRST images are not available for the two sources, namely, J021452.29+140527.4 and J123431.72+645556.5. 
We retain these in our sample.} 
(White et al. 1997). 
Eight sources in the list (J024534.05+010813.8, 
J095631.05+404628.2, J095908.31+024309.6,  J103838.82+494736.8, J130907.98+522437.3, J141701.94+541340.6, 
J160846.76+374850.6 and J223334.89-075043.2) are resolved in the FIRST survey maps 
and the associated radio component closest to the optical source 
has a flux density less than 50 mJy. For this reason these sources are not included in our sample. 
In the case of J124538.35+551132.66 and J144553.46+034732.4 the radio peak of interest is more than  
8$^{\prime\prime}$ away from the corresponding optical positions. As these separations 
are much larger than the astrometric accuracy of FIRST and SDSS we do not include these sources 
in our list. 
Another source, J100121.07+555355.8, is a gravitational lens commonly known as Q0957+561A,B. 
The two optical images are separated by $\sim6^{\prime\prime}$. The radio structure is complex with
core and jets. At 1420 MHz, the radio image is deconvolved into three components and  
none of the peaks coincide with any of the two optical images. The dominant radio component
is located more than 4$^{\prime\prime}$ from the optical sources. For these reasons we did not
include this source in our sample. Note  that 21-cm absorption associated with the
\zabs~=~1.3911 Mg~{\sc ii} system has been searched for at GMRT by Kanekar \& Chengalur (2003). 

We ended up with a complete sample of 34 \mgii (W$_{\rm r}$$\ge$1\AA) systems (in front of 33 radio sources) 
with flux density above 50\,mJy from SDSS DR3.
In our survey we have searched for 21-cm absorption in 24 of these systems. 
The distribution of the radio flux density in these sources is given in the third panel 
of Fig.~\ref{totsample}. From 
this figure it is clear that we have searched for 21-cm absorption in all but the 
two systems with background sources having flux density greater than
100\,mJy. We are only $\sim$45\% complete for flux densities in the range 50$-$100 mJy. 

Using an automatic procedure (similar to the one used by P06), 
we have identified 1548 additional strong 
\mgii systems in the same redshift range from DR5 of SDSS. 
Following the steps discussed above we have identified 29 systems in front of
29 QSOs having at least a S$_{\rm 1.4GHz}>50$\,mJy component 
coinciding with the optical position in the FIRST image. 
Due to scheduling constraints, we could only cover 8 of these sources in our 
GMRT observations. 
%

Therefore in total we have observed a sample of 32 $W_{\rm r}\gapp$1\AA~ Mg~{\sc ii} 
systems in front of the 31 radio sources drawn from SDSS.
In the bottom panel of Fig~\ref{totsample} we give the distribution
of fluxes in our observed sample. 
In summary, we have observed, respectively, $\sim$70\% (resp. 35\%)
of the available sources with flux densities $S_{\rm 1.4GHz}>$~100~mJy
(resp. in the range 50-100~mJy).
From the top two panels of Fig~\ref{totsample} it is apparent that 
the distribution of $W_{\rm r}$(Mg~{\sc ii}$\lambda$2796) and 
Mg~{\sc ii} doublet ratio (DR) in the systems observed with GMRT (filled
circles) represent the parent distribution well.
The two distribution Kolmogorov-Smirnov test (KS test) gives the probability 
of 70\% and 36\% respectively for the distribution of
$W_{\rm r}$(Mg~{\sc ii}$\lambda$2796) and DR(Mg~{\sc ii}) in the sources with
GMRT observations to represent the distribution of the parent population. 
{\it Therefore, despite incompleteness in our survey towards
low flux density levels (i.e $<$100 mJy), the sample is 
representative of the parent population 
as far as} $W_{\rm r}$(Mg~{\sc ii}$\lambda2796$) {\sl and} DR(Mg~{\sc ii})
{\sl are concerned}.

\begin{table*}
\caption{Sample of Mg~II systems observed with GMRT.}
\begin{center}
\begin{tabular}{|l|l|l|c|c|c|c|l|l||l}
\hline
\hline
Source name &   \zem   & \zabs  &  $W_{\rm r}$(\mgiia)  & $W_{\rm r}$(\mgiib)  & $W_{\rm r}$(\mgia) & $W_{\rm r}$(\feiia) & Flux & Morp. & Ref.\\
            &          &        &        (\AA)       &       (\AA)      &     (\AA)      &      (\AA)      &   (mJy)    &      \\
~~~~~~~~(1) & (2)      &  (3)   &  (4)            &   (5)           &     (6)        &     (7)        &    ~~(8)     &  (9)  & (10) \\
\hline
J010826.84-003724.1 & 1.373 & 1.3710& 0.43$\pm$0.05  & 0.39$\pm$0.05  &   0.20$\pm$0.05 &  0.33$\pm$0.06   & 894.9 & C  & Y \\
J015454.36-000723.2 & 1.828 & 1.1803& 1.37$\pm$0.19  & 1.23$\pm$0.19  &$<$0.22          &  0.65$\pm$0.26   & 245.8 & C  & P \\
J021452.29+140527.4 & 2.178 & 1.4463& 2.53$\pm$0.22  & 1.55$\pm$0.19  &   1.04$\pm$0.23 &  1.28$\pm$0.23   & 119.5 & CN & P \\
J0240-2309$^\dag$   & 2.223 & 1.3647& 1.85$\pm$0.002 & 1.65$\pm$0.002 &   0.27$\pm$0.002&  0.87$\pm$0.003  & 6027  & C$^\ddag$  & L\\
J025928.51-001959.9 & 2.000 & 1.3370& 1.76$\pm$0.04  & 1.35$\pm$0.05  &$<$0.05          &  0.36$\pm$0.05   & 226.7 & C  & P \\
J074237.38+394435.6 & 2.200 & 1.1485& 2.48$\pm$0.41  & 1.41$\pm$0.35  &$<$0.26          &$<$0.69           &  99.3 & C  & P \\
J074809.46+300630.5 & 1.695 & 1.4470& 3.68$\pm$0.06  & 3.28$\pm$0.07  &   0.85$\pm$0.08 &   2.17$\pm$0.08  & 200.4 & C  & P \\
J080248.43+291734.1 & 2.380 & 1.3648& 1.04$\pm$0.19  & 0.70$\pm$0.16  &$<$0.79          &$<$0.23           & 190.4 & C  & P \\
J080442.23+301237.0 & 1.452 & 1.1908& 1.28$\pm$0.05  & 1.07$\pm$0.06  &   0.19$\pm$0.06 &   1.05$\pm$0.07  & 1040  & R  & P \\
J080839.66+495036.5 & 1.436 & 1.4071& 1.33$\pm$0.09  & 1.24$\pm$0.08  &   0.48$\pm$0.12 &   0.60$\pm$0.12  & 846.8 & C  & T \\
J081247.78+252242.0 & 1.803 & 1.3683& 3.18$\pm$0.10  & 3.11$\pm$0.11  &   0.87$\pm$0.12 &   2.22$\pm$0.14  & 69.6  & C? & P \\
J084506.24+425718.3 & 2.095 & 1.1147& 1.24$\pm$0.09  & 1.34$\pm$0.09  &   $<$0.11       &   0.87$\pm$0.09  & 203.4 & C  & P \\
J085042.24+515911.7 & 1.894 & 1.3265& 4.56$\pm$0.12  & 4.02$\pm$0.12  &   2.19$\pm$0.12 &   3.44$\pm$0.15  & 61.2  & C  & P \\
J085244.74+343540.4 & 1.655 & 1.3095& 2.91$\pm$0.11  & 2.83$\pm$0.11  &   1.27$\pm$0.12 &   2.26$\pm$0.18  & 66.7  & C  & T \\
J095327.95+322551.6 & 1.575 & 1.2372& 1.55$\pm$0.05  & 1.49$\pm$0.05  &   0.35$\pm$0.06 &   1.18$\pm$0.08  & 127.4 & C  & T \\
J100842.71+621955.7 & 1.875 & 1.2080& 4.04$\pm$0.22  & 3.03$\pm$0.20  &   1.37$\pm$0.30 &   2.38$\pm$0.28  & 73.8  & R  & P \\
J101742.63+535635.0$^*$ & 1.397 & 1.3055& 2.68$\pm$0.10  & 2.62$\pm$0.11  &   0.70$\pm$0.11 &   1.98$\pm$0.14  & 64.0  & R & P \\
J102258.41+123429.7 & 1.727 & 1.2505& 3.06$\pm$0.07  & 2.77$\pm$0.08  &   0.87$\pm$0.07 &   2.45$\pm$0.11  & 64.6  & T  & T \\
J105813.05+493936.1 & 2.396 & 1.2120& 1.91$\pm$0.23  & 1.53$\pm$0.25  &   0.84$\pm$0.20 &   0.88$\pm$0.34  & 215.6 & C  & P \\
J112657.65+451606.3 & 1.811 & 1.3022& 1.26$\pm$0.05  & 1.03$\pm$0.05  &   0.20$\pm$0.05 &   0.38$\pm$0.05  & 374.7 & C  & T \\
J114521.32+045526.7 & 1.342 & 1.3433& 2.15$\pm$0.11  & 1.61$\pm$0.12  &   1.29$\pm$0.14 &   1.40$\pm$0.17  & 413.1 & R  & P \\
J120854.25+544158.1 & 1.344 & 1.2110& 1.85$\pm$0.25  & 2.08$\pm$0.26  &$<$0.36          &   1.22$\pm$0.35  & 236.1 & C  & P \\
J123256.60+572214.1 & 2.118 & 1.3429& 2.32$\pm$0.19  & 1.39$\pm$0.20  &$<$0.19          &   1.53$\pm$0.18  & 107.6 & C  & P \\
J123431.73+645556.5 & 3.036 & 1.3739& 0.95$\pm$0.07  & 0.84$\pm$0.07  &$<$0.08          &   0.63$\pm$0.07  & 88.9  & CN & M \\
                    &       & 1.3829& 1.36$\pm$0.07  & 1.21$\pm$0.07  &$<$0.09          &   0.74$\pm$0.07  &  ''   & '' & P \\
J132901.41+105304.9 & 1.933 & 1.1645& 1.73$\pm$0.18  & 1.51$\pm$0.17  &$<$0.19          &   1.11$\pm$0.17  & 106.9 & C? & T \\
J141104.25-030043.3 & 1.531 & 1.4160& 2.08$\pm$0.11  & 1.74$\pm$0.12  &   0.69$\pm$0.13 &   1.66$\pm$0.13  & 163.0 & R  & P \\
J145633.42+000555.5 & 1.835 & 1.3512& 3.21$\pm$0.21  & 2.02$\pm$0.19  &$<$0.20          &   1.46$\pm$0.26  & 77.8  & R  & P \\
J150823.71+334700.7 & 2.208 & 1.1650& 2.67$\pm$0.08  & 2.93$\pm$0.07  &   0.79$\pm$0.09 &   1.59$\pm$0.12  & 130.3 & C  & T \\
J151005.88+595853.3 & 1.720 & 1.3720& 1.42$\pm$0.07  & 1.21$\pm$0.07  &$<$0.07          &   0.69$\pm$0.10  & 181.0 & D  & P \\
J160456.14-001907.1 & 1.629 & 1.3245& 0.67$\pm$0.05  & 0.68$\pm$0.05  &$<$0.05          &   0.51$\pm$0.05  & 301.6 & T  & L \\
J162346.23+071854.9 & 1.648 & 1.3350& 0.93$\pm$0.06  & 0.91$\pm$0.06  &   0.13$\pm$0.05 &   0.68$\pm$0.06  & 69.3  & C  & T \\  
J231222.36-010924.8 & 1.431 & 1.4262& 2.73$\pm$0.24  & 1.84$\pm$0.24  &   0.46$\pm$0.45 &   1.25$\pm$0.60  & 237.7 & R  & P \\ 
J234023.66-005326.9 & 2.085 & 1.3603& 1.60$\pm$0.04  & 1.59$\pm$0.04  &   0.50$\pm$0.04 &   1.18$\pm$0.04  & 114.2 & C  & P \\
J235810.87-102008.7 & 1.638 & 1.1726& 1.52$\pm$0.15  & 1.81$\pm$0.14  &   0.50$\pm$0.14 &   1.11$\pm$0.16  & 730.7 & C  & P \\  
\hline
\end{tabular}
\end{center}
\begin{flushleft}
Col. 1: SDSS source name; col. 2: QSO emission redshift; col. 3: Absorption redshift of \mgii systems as determined from \mgiiab, 
\mgia and \feiia absorption lines; 
cols. 4, 5, 6, 7: Rest equivalent widths or 1$\sigma$ limit of \mgiia, \mgiib, \mgia, \feiia~absorption lines respectively;  
col. 8: Peak flux density at 1.4 GHz as measured from FIRST (or NVSS when FIRST image is not available) survey, 
col. 9: Radio morphology as determined from FIRST (or NVSS) where C=compact at the FIRST resolution ($\sim$5$^{\prime\prime}$), 
CN=compact at NVSS resolution ($\sim$45$^{\prime\prime}$), R=resolved in the FIRST image, D=double lobed and T=triple radio 
source, and col. 10: Origin of the Mg~{\sc ii} systems, where Y is for York et al. (2006), P is for 
P06, L for Lanzetta et al. (1987), T for `this paper' i.e. drawn from SDSS using a procedure 
developed by one of us and M is for detected while manually checking the SDSS spectra for other systems.  \\   
$^{\dag}$ Absorption line equivalent widths for this system are estimated from the VLT UVES data presented in Srianand et al. (2007). \\ 
$^{\ddag}$ Source is not covered by the FIRST and NVSS surveys but is a VLA phase calibrator for A, B, C and D configurations. \\
$^*$ This source is extended in FIRST but compact in our GMRT image.\\
\end{flushleft}
\label{mg2sample}
\end{table*}


In addition, we have observed three more sources i.e.
J010826.84$-$003724.1 (York et al. 2006), and J0240-2309 and J160456.14$-$001907.1 
(Lanzetta et al. 1987) because they have a total flux density at 610\,MHz 
in excess of 1 Jy and are associated with strong Mg~{\sc ii}
absorptions. We notice that addition of these systems (that were 
observed in the early stages of our survey) does not change the
statistical conclusions drawn above.
Thus, in total we have observed 35 \mgii systems, 33 of which have $W_{\rm r}$(Mg~{\sc ii})$\gapp$ 1\AA, 
to search for 21-cm absorption. In Table~\ref{mg2sample} we give the list of observed sources 
together with the emission redshift, \zem in column 2, the absorption redshift (\zabs) in column 3, 
the rest equivalent widths of Mg~{\sc ii}$\lambda$$\lambda$2796,2803 (columns 4 and 5), 
Mg~{\sc i}$\lambda$2852 (column 6) and Fe~{\sc ii}$\lambda$2600 (column 7), 
the radio flux density at 1420 MHz (column 8), morphology of the radio sources (column 9) and
the origin of the Mg~{\sc ii} systems (column 10).
The equivalent width estimates in Table~\ref{mg2sample} were 
obtained by integrating the absorption over the absorption profile. 
For consistency these measurements were compared against estimates from Voigt profile 
fits. Differences of the order of 0.2\AA~ or less were found 
between estimates from these methods. This is similar to uncertainties ($\sim$0.3\AA) 
from estimating equivalent widths by automatic procedures (see P06).

Preliminary results from the first phase of the GMRT survey reporting 
3 new detections in 10 systems were presented in Gupta et al. 2007 
(see also Srianand, Gupta \& Petitjean, 2007 for details regarding J0240$-$2309). 
Detailed analysis of two detections from our
survey with strong signatures of dust reddening is presented in Srianand 
et al. (2008b). Here we present the results for the entire survey.

\section{GMRT Observations}

\begin{table}
\caption{Observing log for the GMRT observations.}
\begin{center}
\begin{tabular}{|l|l|l|l|l|}
\hline
\hline
Source name &   \zabs  &Date & Time  & Ch. Width  \\
            &          &     & (hr)  &  (km\,s$^{-1}$) \\
~~~~~~~~(1) &    (2)   & (3) &  (4)  & (5) \\
\hline
J0108$-$0037 &  1.3710  & 2007 Jul 03        &4.4     &  3.9    \\
J0154$-$0007 &  1.1803  & 2007 Jan 30        &5.3     &  3.6    \\
J0214+1405   &  1.4463  & 2006 Jul 16        &4.4     &  4.0    \\
J0240$-$2309 &  1.3647  & 2004 Feb 09        &8.0     &  3.9    \\
J0259$-$0019 &  1.3370  & 2006 Oct 06        &2.3     &  3.9    \\
J0742+3944   &  1.1485  & 2007 Dec 31        &6.3     &  3.5    \\
J0748+3006   &  1.4470  & 2006 Aug 01        &5.5     &  4.0    \\
J0802+2917   &  1.3648  & 2007 Nov 26        &7.0     &  3.9    \\
J0804+3012   &  1.1908  & 2006 Aug 22        &6.0     &  3.6    \\
             &          & 2006 Sep 15,16     &6.2     &  7.2$^{2,128}$    \\
J0808+4950   &  1.4071  & 2006 Dec 06        &5.3     &  4.0    \\
             &          & 2008 May 08        &5.1     &  4.0    \\
J0812+2522   &  1.3683  & 2007 Feb 01        &5.8     &  3.9    \\
J0845+4257   &  1.1147  & 2006 Sep 09        &5.0     &  3.5    \\
J0850+5159   &  1.3265  & 2007 Nov 06        &7.3     &  3.8    \\
J0852+3435   &  1.3095  & 2007 Nov 05        &7.5     &  3.8    \\
             &          & 2007 Nov 30        &6.9     &  3.8    \\
             &          & 2008 Mar 08        &6.2     &  3.8    \\
J0953+3225   &  1.2372  & 2007 Jun 05        &5.8     &  3.7    \\
J1008+6219   &  1.2080  & 2006 Oct 06        &6.5     &  3.6    \\
J1017+5356   &  1.3055  & 2006 Aug 29        &4.4     &  3.8    \\
J1022+1234   &  1.2505  & 2007 Nov 29        &8.3     &  3.7    \\
J1058+4939   &  1.2120  & 2008 May 10        &5.9     &  3.6    \\
J1126+4516   &  1.3022  & 2007 Dec 29        &5.5     &  3.8    \\
J1145+0455   &  1.3433  & 2007 Jan 30        &5.2     &  3.9    \\
J1208+5441   &  1.2110  & 2008 May 11        &6.1     &  3.7    \\
J1232+5722   &  1.3429  & 2007 Dec 28        &5.6     &  3.9    \\
J1234+6455   &  1.3739  & 2008 Jan 02        &5.6     &  3.9    \\
             &  1.3829  & 2007 Dec 30        &5.4     &  3.9    \\
J1329+1053   &  1.1645  & 2007 Jun 21        &5.2     &  3.6    \\
J1411$-$0300 &  1.4160  & 2006 Jul 06,07   &3.9     &  4.0    \\
J1456+0005   &  1.3512  & 2008 Jan 01        &6.1     &  3.9    \\
J1508+3347   &  1.1650  & 2007 Jun 05        &4.1     &  3.6    \\
J1510+5958   &  1.3720  & 2007 Feb 02        &5.6     &  3.9    \\
J1604$-$0019 &  1.3245  & 2006 Jun 19        &3.3     &  3.8$^{2,256}$    \\
J1623+0718   &  1.3350  & 2007 Dec 16        &5.4     &  3.9    \\
             &          & 2008 May 09        &6.3     &  3.9    \\
J2312$-$0109 &  1.4262  & 2006 Oct 04        &4.3     &  4.0    \\
J2340$-$0053 &  1.3603  & 2007 Jan 21        &5.8     &  3.9    \\
             &          & 2007 Apr 03        &5.8     &  1.0$^{0.25,128}$    \\
J2358$-$1020 &  1.1726  & 2006 May 23        &3.9     &  3.6    \\  
             &          & 2006 Jul 30        &5.5     &  1.8$^{0.5,128}$    \\  
\hline
\end{tabular}
\end{center}
\begin{flushleft}
Col. 1: Source name; col. 2: \mgii absorption redshift; 
col. 3: Date of observation; col. 4: Time on source; col. 5: Channel width in km\,s$^{-1}$. 
Superscripts to channel width give the baseband bandwidth in MHz and number of spectra channels for 
the observation when a baseband set-up different from 1\,MHz split into 128 channels was used.  \\ 
\end{flushleft}
\label{mg2obslog}
\end{table}


\begin{figure*}
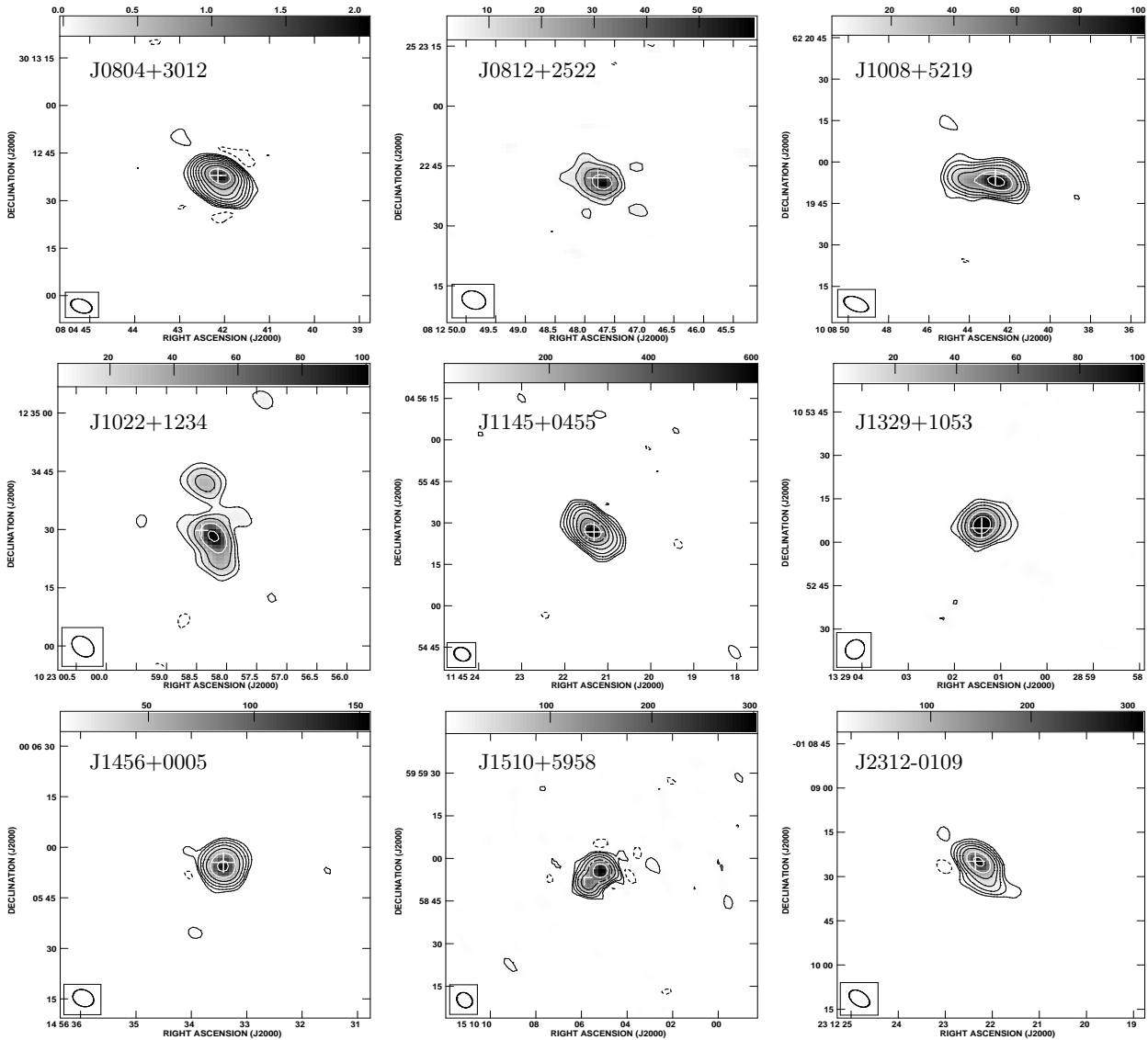

\centerline{
\vbox{
\hbox{
\psfig{figure=j0804_map.ps,height=5.0cm,width=5.5cm,angle=-90}
\psfig{figure=j0812_map.ps,height=5.0cm,width=5.5cm,angle=-90}
\psfig{figure=j1008_map.ps,height=5.0cm,width=5.5cm,angle=-90}
}
\hbox{ 
\psfig{figure=j1022_map.ps,height=5.0cm,width=5.5cm,angle=-90}
\psfig{figure=j1145_map.ps,height=5.0cm,width=5.5cm,angle=-90}
\psfig{figure=j1329_map.ps,height=5.0cm,width=5.5cm,angle=-90}
}
\hbox{
\psfig{figure=j1456_map.ps,height=5.0cm,width=5.5cm,angle=-90}
\psfig{figure=j1510_map.ps,height=5.0cm,width=5.5cm,angle=-90}
\psfig{figure=j2312_map.ps,height=5.0cm,width=5.5cm,angle=-90}
}
}
}
\caption[]{GMRT maps of sources resolved in our observations with contour
levels as
n$\times$(-1,1,2,4, ...) mJy beam$^{-1}$.  
J0804+3012: 0.84 mJy beam$^{-1}$ rms and beam of
6.36$^{\prime\prime}$$\times$4.09$^{\prime\prime}$,
position angle PA=67$^{\circ}$ and n=4.0.
J0812+2522: 0.90 mJy beam$^{-1}$ rms and beam of
5.60$^{\prime\prime}$$\times$4.37$^{\prime\prime}$,
PA=66$^{\circ}$ and n=5.0.
J1008+5219: 0.51 mJy beam$^{-1}$ rms and beam of
8.87$^{\prime\prime}$$\times$4.92$^{\prime\prime}$,
P.A.=67$^{\circ}$ and n=1.6.
J1022+1234: 0.90 mJy beam$^{-1}$ rms and beam of
6.13$^{\prime\prime}$$\times$4.49$^{\prime\prime}$,
PA=45$^{\circ}$ and n=6.0.
J1145+0455: 0.90 mJy beam$^{-1}$ rms and beam of
5.85$^{\prime\prime}$$\times$4.70$^{\prime\prime}$,
PA=63$^{\circ}$ and n=3.0.
J1329+1053: 0.70 mJy beam$^{-1}$ rms and beam of
6.85$^{\prime\prime}$$\times$5.88$^{\prime\prime}$,
PA=$-$26$^{\circ}$ and n=3.2.
J1456+0005: 0.48 mJy beam$^{-1}$ rms and beam of
5.93$^{\prime\prime}$$\times$4.78$^{\prime\prime}$,
PA=59$^{\circ}$ and n=2.0.
J1510+5958: 1.10 mJy beam$^{-1}$ rms and beam of
5.67$^{\prime\prime}$$\times$4.71$^{\prime\prime}$,
PA=36$^{\circ}$ and n=6.0.
J2312$-$0109: 0.95 mJy beam$^{-1}$ rms and beam of
7.38$^{\prime\prime}$$\times$4.66$^{\prime\prime}$,
PA=53$^{\circ}$ and n=4.0.
}
\vskip -18.0cm
\begin{picture}(400,400)(0,0)
%
\put(000,370){\normalsize J0804+3012}
\put(160,370){\normalsize J0812+2522}
\put(315,370){\normalsize J1008+5219}
\put(000,225){\normalsize J1022+1234}
\put(160,225){\normalsize J1145+0455}
\put(315,225){\normalsize J1329+1053}
\put(000,085){\normalsize J1456+0005}
\put(160,085){\normalsize J1510+5958}
\put(315,085){\normalsize J2312-0109}
\end{picture}
\vskip +4.0cm
\label{gmrtmap}
\end{figure*}

\begin{table*}
\caption{Observational results for \mgii systems observed with GMRT.}
\begin{center}
\begin{tabular}{|l|l|l|l|l|l|r|l|r|}
\hline
\hline
Source name &   \zabs  &  Morph. & Peak Flux  & $\delta v$ & Spectral rms & $\tau$ & $\tau_{3\sigma,10}$ & $\int\tau$dv\\
            &          &         & (mJy)      &  (km\,s$^{-1}$) &  (mJy\,b$^{-1}$\,ch$^{-1}$)  &        &          & (km\,s$^{-1}$) \\
~~~~~~~~(1) &    (2)   & (3) &  (4)  &  (5)    &     (6)  &     (7)   &  (8)  & (9)   \\
\hline
J0108$-$0037 &  1.3710  &   C    & 1276     & 3.9   &     2.9    &    0.070  & 0.003 &    1.29  \\  
J0154$-$0007 &  1.1803  &   C    & 154      & 3.6   &     1.7    & $<$0.011  & 0.016 & $<$0.17  \\  
J0214+1405   &  1.4463  &   C    & 220      & 4.0   &     2.7    & $<$0.012  & 0.025 & $<$0.26  \\  
J0240$-$2309 &  1.3647  &   C    & 5100     & 3.9   &     5.2    & $<$0.001  & 0.002 & $<$0.02  \\  
J0259$-$0019 &  1.3370  &   C    & 149      & 3.9   &     2.4    & $<$0.016  & 0.027 & $<$0.28  \\  
J0742+3944   &  1.1485  &   C    & 152      & 3.5   &     1.9    & $<$0.013  & 0.034 & $<$0.36  \\  
J0748+3006   &  1.4470  &   C    & 347      & 4.0   &     3.5    & $<$0.010  & 0.017 & $<$0.18  \\  
J0802+2917   &  1.3648  &   C    & 404      & 3.9   &     1.3    & $<$0.003  & 0.006 & $<$0.06  \\  
J0804+3012   &  1.1908  &   R    & 2069     & 7.2   &     1.5    &    0.006  & 0.002 &    0.39  \\  
J0808+4950   &  1.4071  &   C    & 754      & 4.0   &     1.2    &    0.008  & 0.002 &    0.10  \\  
J0812+2522   &  1.3683  &   R    & 59.1     & 3.9   &     1.2    & $<$0.020  & 0.036 & $<$0.38  \\  
J0845+4257   &  1.1147  &   C    & 224      & 3.5   &     3.0    & $<$0.013  & 0.026 & $<$0.27  \\  
J0850+5159   &  1.3265  &   C    & 64.0     & 3.8   &     1.2    &    0.503  & 0.038 &    15.3  \\  
J0852+3435   &  1.3095  &   C    & 51.2     & 3.8   &     0.8    &    0.143  & 0.036 &    6.91  \\  
J0953+3225   &  1.2372  &   C    & 148      & 3.7   &     1.8    & $<$0.015  & 0.020 & $<$0.22  \\  
J1008+6219   &  1.2080  &   R    & 132      & 3.6   &     1.0    & $<$0.008  & 0.014 & $<$0.15  \\  
J1017+5356   &  1.3055  &   R    & 127      & 3.8   &     2.3    & $<$0.018  & 0.036 & $<$0.38  \\  
J1022+1234   &  1.2505  &   T    & 109      & 3.7   &     1.2    & $<$0.011  & 0.020 & $<$0.21  \\  
J1058+4939   &  1.2120  &   C    & 437      & 3.6   &     1.3    &    0.020  & 0.005 &    0.41  \\  
J1126+4516   &  1.3022  &   C    & 356      & 3.8   &     1.4    & $<$0.004  & 0.008 & $<$0.08  \\  
J1145+0455   &  1.3433  &   R    & 717      & 3.9   &     2.4    & $<$0.003  & 0.006 & $<$0.07  \\  
J1208+5441   &  1.2110  &   C    & 274      & 3.7   &     1.5    & $<$0.005  & 0.009 & $<$0.10  \\  
J1232+5722   &  1.3429  &   C    & 221      & 3.9   &     1.4    & $<$0.007  & 0.011 & $<$0.12  \\  
J1234+6455   &  1.3739  &   C    & 141      & 3.9   &     1.3    & $<$0.009  & 0.017 & $<$0.18  \\  
             &  1.3829  &   C    & 139      & 3.9   &     1.4    & $<$0.010  & 0.017 & $<$0.18  \\  
J1329+1053   &  1.1645  &   R    & 171      & 3.6   &     2.6    & $<$0.015  & 0.034 & $<$0.36  \\  
J1411$-$0300 &  1.4160  &   R    & 244      & 4.0   &     3.0    & $<$0.012  & 0.026 & $<$0.28  \\  
J1456+0005   &  1.3512  &   R    & 152      & 3.9   &     1.0    & $<$0.007  & 0.014 & $<$0.15  \\  
J1508+3347   &  1.1650  &   C    & 219      & 3.6   &     2.4    & $<$0.007  & 0.022 & $<$0.23  \\  
J1510+5958   &  1.3720  &   D    & 173      & 3.9   &     1.4    & $<$0.008  & 0.016 & $<$0.17  \\  
J1604$-$0019 &  1.3245  &   T    & 375      & 3.8   &     3.9    & $<$0.010  & 0.019 & $<$0.20  \\  
J1623+0718   &  1.3350  &   C    & 142      & 3.9   &     1.0    &    0.040  & 0.013 &    0.91  \\  
J2312$-$0109 &  1.4262  &   R    & 310      & 4.0   &     2.4    & $<$0.008  & 0.015 & $<$0.16  \\  
J2340$-$0053 &  1.3603  &   C    & 43.4     & 1.0   &     2.3    &    0.754  & 0.042 &    2.50  \\  
J2358$-$1020 &  1.1726  &   C    & 432      & 1.8   &     2.5    &    0.035  & 0.007 &    0.23  \\  
\hline
\end{tabular}
\end{center}
\begin{flushleft}
Col. 1: Source name; col. 2: \mgii absorption redshift; col. 3: Morphology of the radio source as determined from our 
observations; col. 4: Peak flux density in mJy; col. 5 and 6: Spectral resolution in km\,s$^{-1}$ and spectral rms in 
mJy beam$^{-1}$ channel$^{-1}$; col. 7: Maximum of optical depth in one channel or 1$\sigma$ limit to it; 
col. 8: 3$\sigma$ optical depth limit for spectra smoothed to $\sim$10\,km\,s$^{-1}$; col. 9: integrated 21-cm optical depth 
or 3$\sigma$ upper limit in case of non-detections
considering $\delta v$=10\,km\,s$^{-1}$. 
 \\
\end{flushleft}
\label{mg2obsres}
\end{table*}

\begin{figure*}
\centerline{{
\psfig{figure=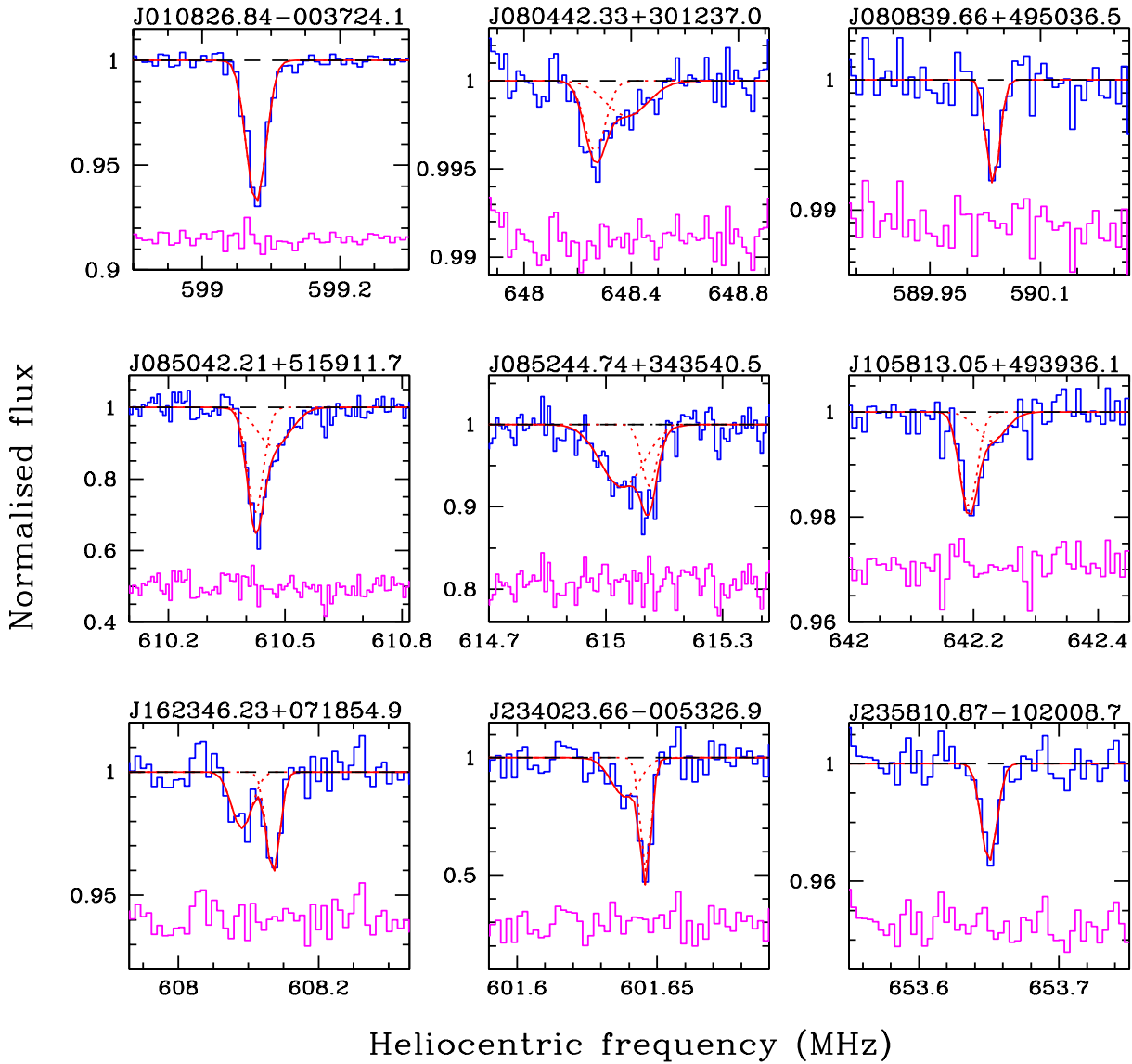,height=16.0cm,width=17.0cm,angle=0}
}}
\caption[]{GMRT spectra of detected 21-cm absorption lines. Individual Gaussian components and  
resultant fits to the absorption profiles are overplotted as dotted and continuous lines respectively.  
Residuals, on an offset arbitrarily shifted for clarity, are also shown.
}
\label{mg2det}
\end{figure*}


\begin{figure*}
\centerline{{
\psfig{figure=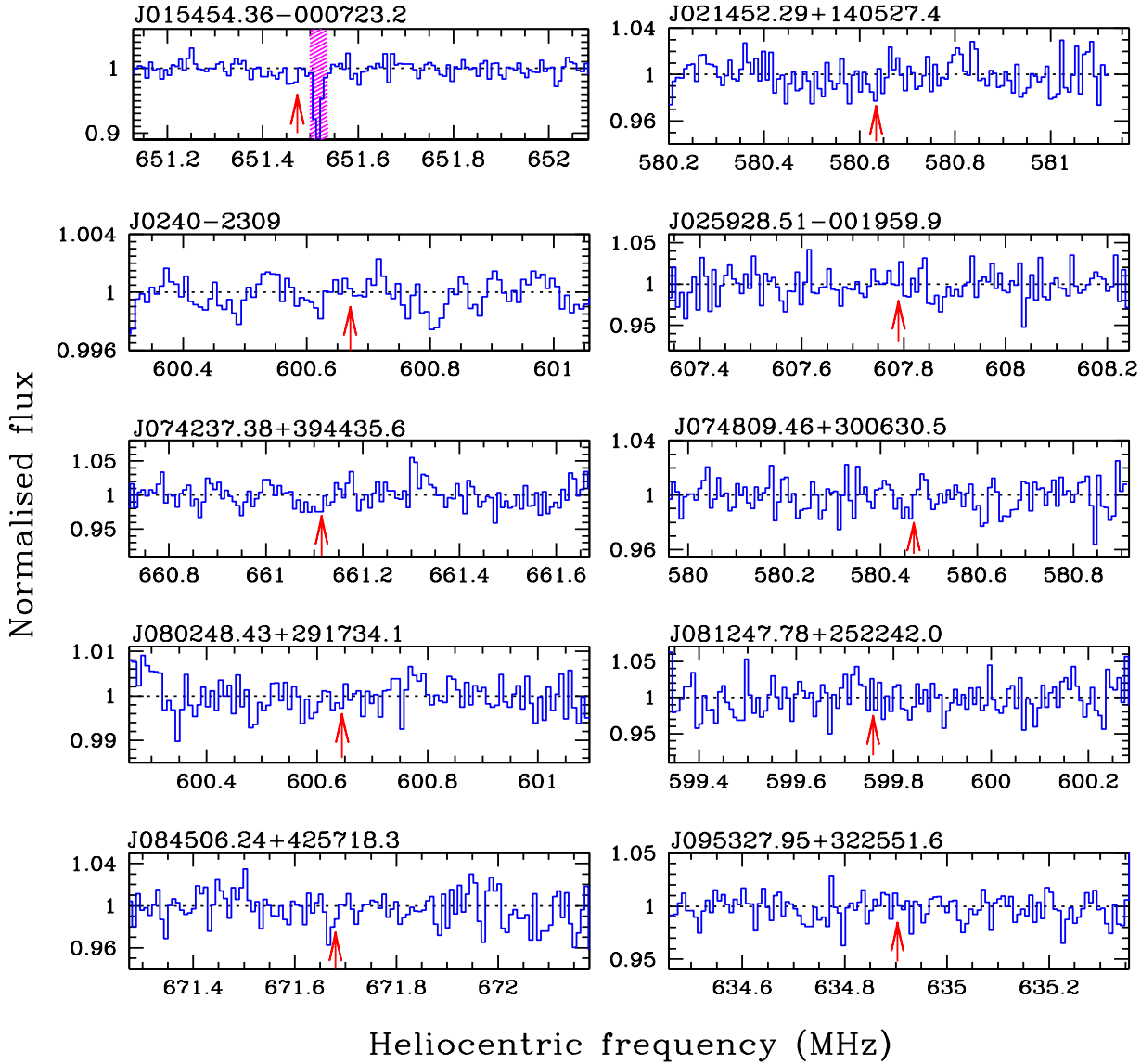,height=16.0cm,width=17.0cm,angle=0}
}}
\caption[]{GMRT spectra of sources with non-detection.  
Arrows mark the expected positions of 21-cm absorption lines based on metal absorption lines.  
The shaded regions mark frequency ranges affected by RFI.
The \zabs=1.3739 and 1.3829 absorbers towards J1234+6455 are referred to as A and B respectively.  
For J1411-0300, P1 and P2 correspond to spectrum towards two peaks (see Gupta et al. 2007 for detail).  
The spectra toward northern and southern radio peaks of J1510+5958 are marked as N and S.
}
\label{mg2nondet1}
\end{figure*}


\begin{figure*}
\addtocounter{figure}{-1}
\centerline{{
\psfig{figure=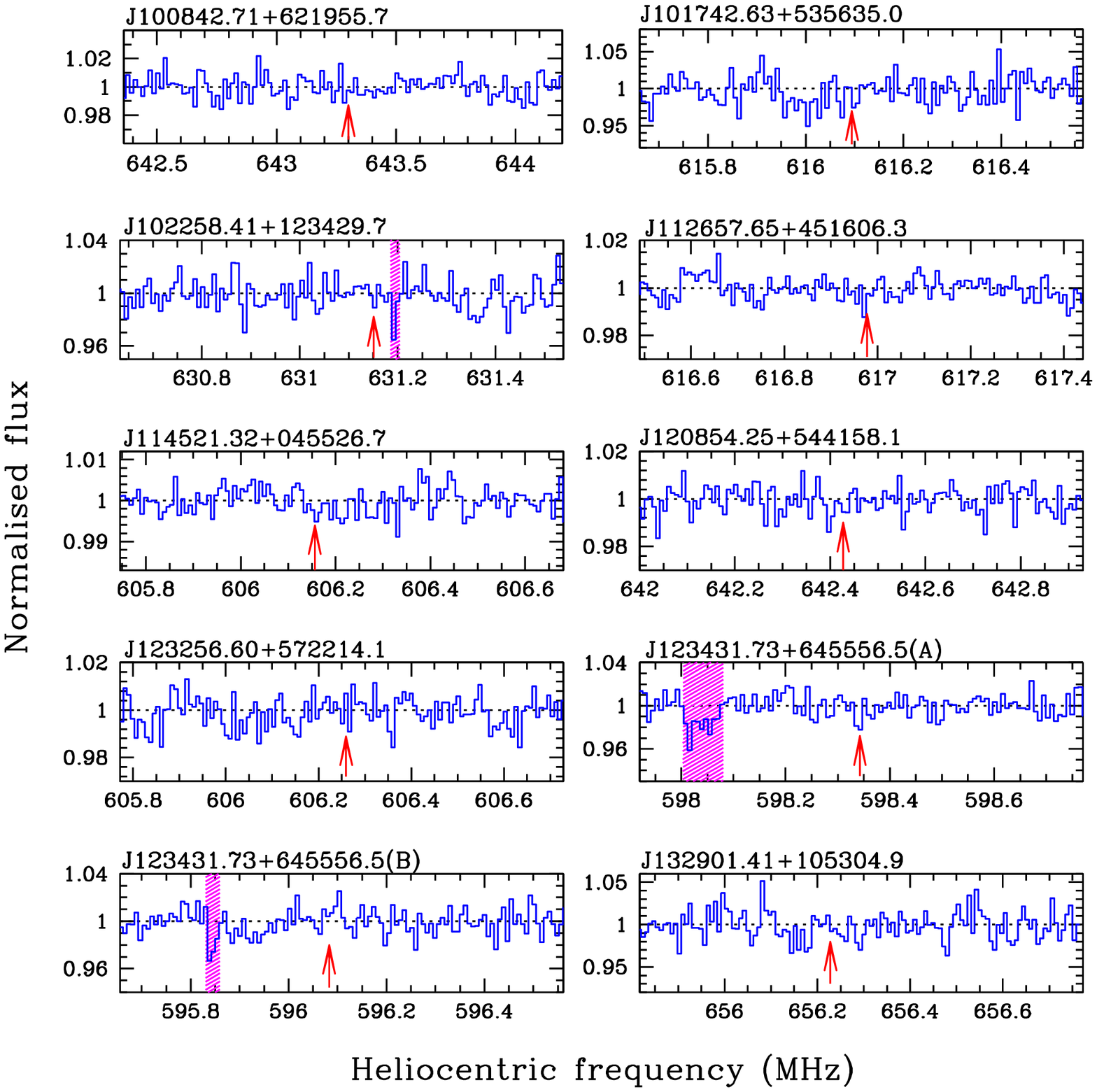,height=16.0cm,width=17.0cm,angle=0}
}}
\caption[]{{\sl Continued}. 
}
\label{mg2nondet2}
\end{figure*}


\begin{figure*}
\addtocounter{figure}{-1}
\centerline{{
\psfig{figure=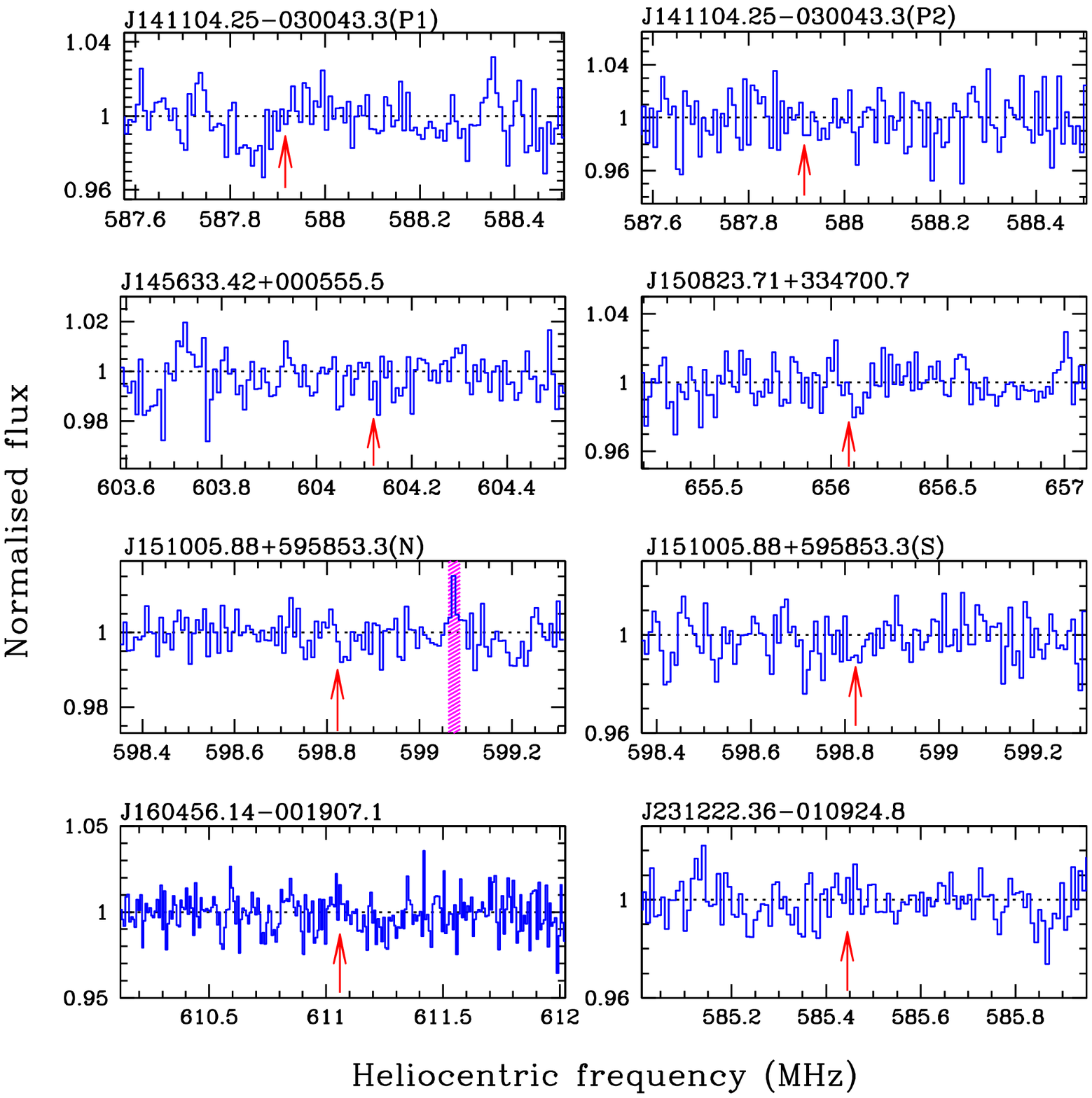,height=13.9cm,width=17.0cm,angle=0}
}}
\caption[]{{\sl Continued}.  
}
\label{mg2nondet2}
\end{figure*}

%
%
%
We observed 35 \mgii systems 
with GMRT 610-MHz band using in total $\sim$400\,hrs of telescope time. 
Details of the observations are given in Table~\ref{mg2obslog}.
For our survey, we have usually used the 1\,MHz baseband bandwidth split into 128 frequency 
channels yielding a spectral resolution of $\sim$4\,km\,s$^{-1}$. 
Broader (2\,MHz for J0804+3012) and narrower (0.5\,MHz for J2358$-$1020 
and 0.25\,MHz for J2340-0053) bandwidths were used as well
for confirming the detected 21-cm absorptions.
During each observations, the local oscillator chain was tuned to centre the 
baseband at the redshifted 21-cm frequency defined by the mean redshift 
estimated from the \mgii and other absorption lines detected in the 
SDSS spectrum. 
Data were acquired in the two circular polarization channels called RR and LL. 
We typically observed each absorption line system for 8\,hrs which included 
calibration overheads. For flux and bandpass calibration we observed one of 
the standard flux density calibrators 3C\,48, 3C\,147 and 3C\,286 for 
10-15\,mins every two hours.  A phase calibrator was also observed for 
10\,mins every $\sim$40\,mins to get reliable phase solutions. Total 
on-source time after excluding the telescope set-up time and calibration 
overheads are provided in Table~\ref{mg2obslog}.    

GMRT data were reduced using the NRAO Astronomical Image Processing System 
(AIPS) following the standard procedures as described in Gupta et al. (2006).  
After the initial flagging and calibration, source and calibrator data were 
examined to flag and exclude the baselines and timestamps affected by Radio 
Frequency Interference (RFI). Applying complex gains and bandpass obtained 
using the flux and phase calibrators, a continuum map of the source was made 
using absorption-free channels. Using this map as a model, self-calibration 
complex gains were determined and applied to all the frequency channels.  The 
same continuum map was then used to subtract the continuum emission from 
the dataset. This continuum subtracted dataset was then imaged separately 
in stokes RR and LL to get 3-dimensional (with third axis as frequency) data 
cubes.  
The spectra at the quasar positions were extracted from these cubes 
and compared for consistency. If necessary a first-order 
cubic-spline was fitted to remove the residual continuum from spectra. 
The two polarization channels were then combined to get the final stokes I 
spectrum, which was then shifted to heliocentric frame.    

We have obtained useful radio spectra for 35 Mg~{\sc ii} 
systems in front of 34 radio sources.  Of these 34 
radio sources 20 are compact in FIRST (deconvolved sizes 
$<2^{\prime\prime}$) and our 610 MHz images.  These are assigned 
morphology C in Tables~\ref{mg2sample} and \ref{mg2obsres}.  
Of the remaining 14 sources, for two sources (i.e. J0214+1405 and J1234+6455) 
FIRST images are not available.    
Both sources are compact in NVSS and our GMRT 610-MHz images.  
Radio emission associated with J0214+1405 in our 
GMRT image (with beam 5.5$^{\prime\prime}\times$4.6$^{\prime\prime}$ and PA=57$^\circ$) 
is well represented by a Gaussian component implying a 
deconvolved size of 2.5$^{\prime\prime}\times$1.6$^{\prime\prime}$ with position 
angle 56$^\circ$.  Despite the deconvolved radio source size 
being slightly greater than 2$^{\prime\prime}$, $>$90\% of the 
total emission is contained in the above mentioned 
Gaussian component, so we consider this source to be compact. 
Gaussian deconvolution of J1234$+$6455 yields a size of 
0.7$^{\prime\prime}\times$0.0$^{\prime\prime}$ with position angle 12$^\circ$ 
implying that the source is compact. This source is also compact in sub-arcsecond 
scale CLASS images at 8.4\,GHz. Thus there are 22 sources compact at arcsec scales
in our sample and 23 Mg~{\sc ii} systems along their lines of sight with GMRT data. 
From Table~\ref{mg2obsres} one can see eight (i.e $\sim$35\%) of these systems 
show detectable 21-cm absorption.  

Eleven sources in our sample show extensions in our GMRT continuum maps 
(see Fig.~\ref{gmrtmap} in this article and Fig. 4 in Gupta et al. 2007). For 8 of these sources
spectra were extracted at the position of the peak flux that also coincides well with 
the optical position. For the three remaining sources (J1604$-$0019, J1411$-$0300 
and J1510+5958) we also extracted spectra from the hot spots that are resolved 
from the core in our GMRT observations. In none of these cases we have detected 21-cm 
absorption towards hot spots. We address the issue of radio morphology of our target 
sources and its influence on the detectability of 21-cm absorption in following sections.
Absorption profiles of the nine newly discovered 21-cm absorbers are presented in 
Fig.~\ref{mg2det} whereas spectra with no detectable 21-cm absorption 
are shown in Fig.~\ref{mg2nondet1}. Results are summarised in Table~\ref{mg2obsres}. 
In the 3rd to 6th columns of this table we  provide the morphology, the
peak flux ($F_{\rm P}$) measured in our GMRT 610 MHz continuum image, the velocity 
resolution ($\delta v$) and the measured spectral rms. In the seventh column,
we provide the maximum optical depth per channel 
or the 1$\sigma$ limit on it. The next column gives the 3$\sigma$ optical depth limit 
obtained as follows: $\tau_{3\sigma} = -ln(1-3\Delta F/F_{\rm P})$, where $F_{\rm P}$ is 
the peak flux and $\Delta F$ is the spectral rms for the spectrum smoothed to 10 \kms resolution. 
This smoothing velocity scale is chosen to enable comparative studies with the existing low redshift 
(i.e \zabs $\le$1.0) measurements (see Section 7). In the case of detections, we measure this 
quantity in the line free spectral range. The last column in the table gives the  
integrated optical depth  or the limit on it. 
In the later case we used a Gaussian width of 10 \kms~ for the integration
(i.e 10.6$\times \tau_{3\sigma}$).

For detections, it is important to rule out the possibility of false identification even in case 
of good redshift coincidence. For spectra with high SNR (i.e 21-cm absorption lines towards
J0108$-$0037, J0850$+$5159 and  J1058$+$4939) 
we split the data into different time and UV ranges and check for consistency of the spectra.
We also compared the spectra with spectra of other bright sources in the field.
For spectra of poor SNR or when the narrow absorption line feature is spread over only
one or two frequency channels  (i.e. along the lines of sight towards
J0804+3012, J0808+4950, J0852+3435, J1623+0718, J2340$-$0053 and J2358$-$1020), observations were 
repeated at a different epoch.  
In all cases we observed a consistent absorption profile at a frequency that is in good agreement with
the Doppler shift expected  from the heliocentric motion. 
For J0808+4950, J0852+3435 and J1623+0718 spectra at different epochs were obtained at the same 
spectral resolution ($\sim$\,4 \kms) and the final spectra presented in Fig.~\ref{mg2det} 
are weighted averages of these. The absorption line system towards J0804+3012 was observed 
three times. The later two observations were done using 2\,MHz bandwidth split into 128 channels 
to obtain a wider spectral coverage. The final spectrum presented here was obtained 
by smoothing the spectrum from the first observing run and combining this 
with the later two spectra of spectral resolution $\sim$7\,\kms (see also Gupta et al. 2007).  
Narrow 21-cm absorption profiles detected towards J2340$-$0053 and J2358$-$1020 
were reobserved at higher spectral resolution.  Here, we present only the highest resolution 
spectra.  

\section{Details of new 21-cm absorbers}
\begin{figure}
\psfig{figure=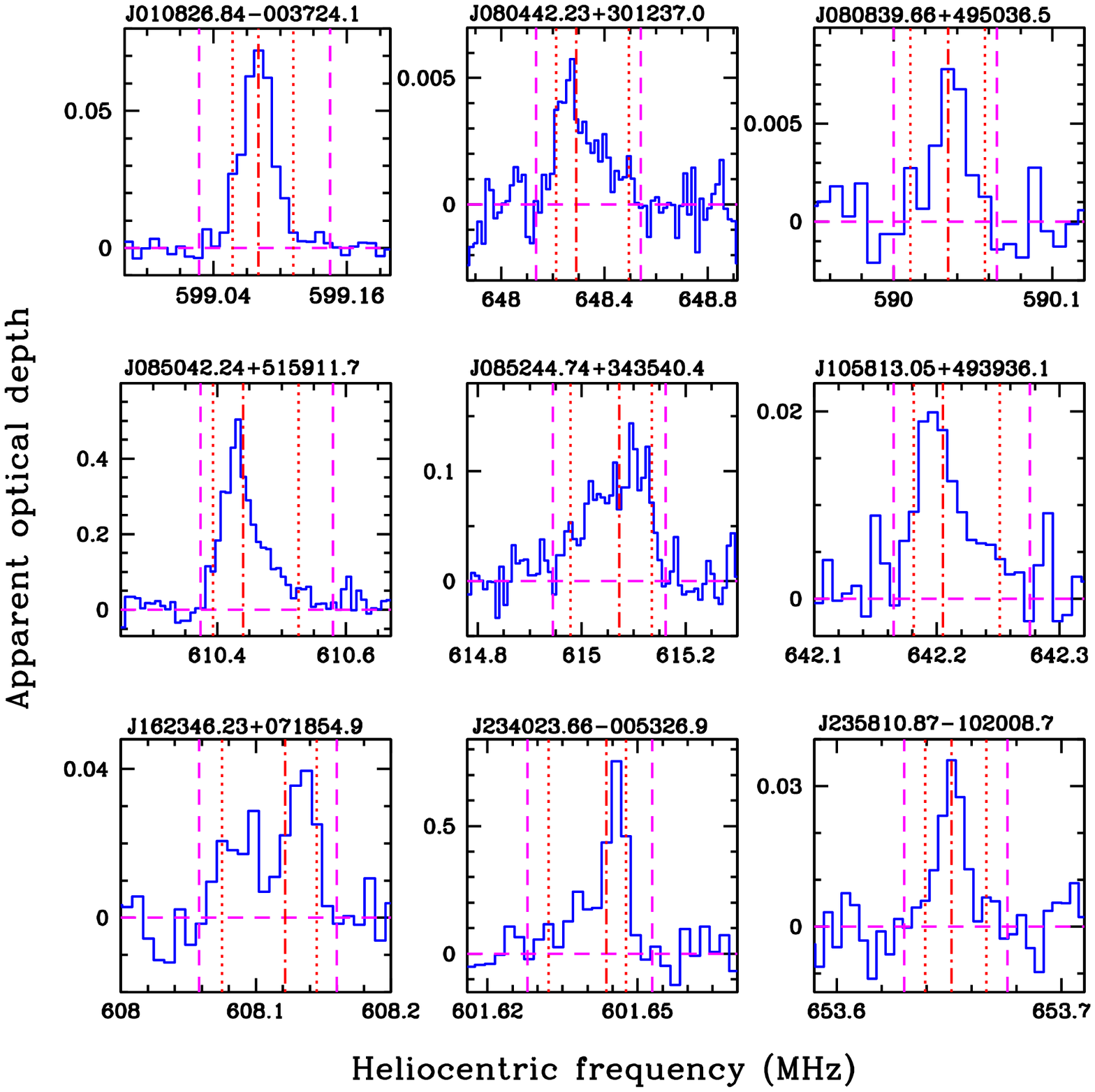,height=9.0cm,width=9.0cm,angle=0}
\caption{Apparent optical depth of 21-cm absorption features detected
in our GMRT survey. The outermost dashed vertical lines give the velocity range 
over which the optical depth is integrated. The dotted vertical lines 
correspond to the width of the profile minus 5\% of the absorption
on both sides (see Fig. 1 of Ledoux et al. 2006).
}
\label{aop}
\end{figure}
\begin{figure*}
\centerline{{
\psfig{figure=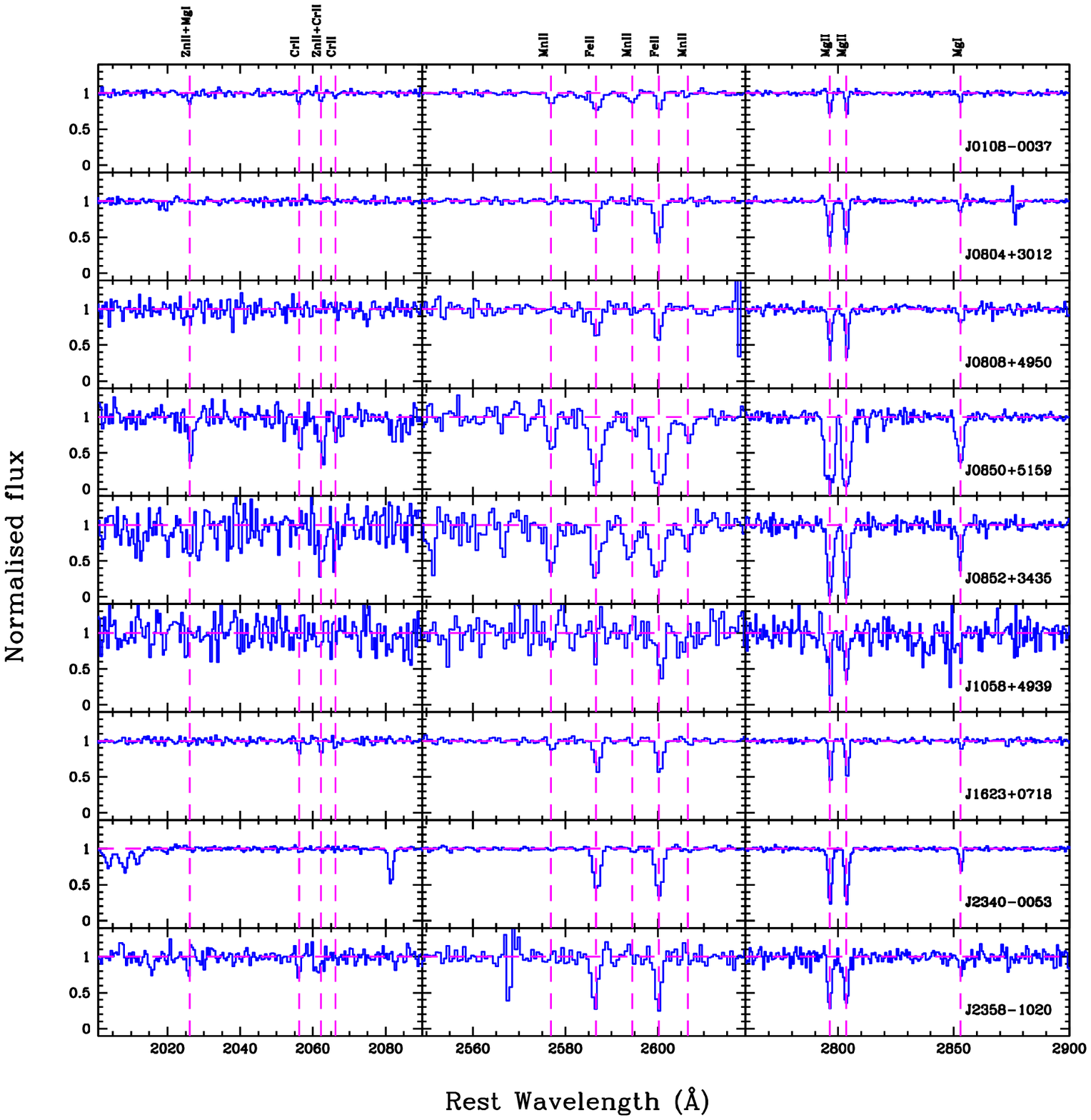,height=16.0cm,width=17.0cm,angle=0}
}}
\caption[]{Normalised rest frame SDSS spectra of the \mgii systems 
with 21-cm detection. Vertical dashed lines mark the location of interesting 
metal line transitions.}
\label{metalfig}
\end{figure*}

\begin{table*}
\caption{Characteristics of 21-cm absorption profiles and multiple Gaussian fits.}
\begin{tabular*}{1.0\textwidth}{@{\extracolsep{\fill}}lccccccc}
\hline
\hline
QSO &$\int\tau dv$ & $\Delta V$ & ${f_{\rm c}N({\rm HI})^a}\over T_{\rm s}$  &$z_{\rm abs}$ & $\Delta v^b$ & $\tau^c_{\rm p}$ &  ${f_{\rm c}N({\rm HI})^d\over T_{\rm s}}$ \\
\hline
J0108$-$0037 &1.29$\pm$0.04& 27.31&2.37$\pm$0.07  &1.37099    &  16$\pm$1    & 0.070$\pm$0.002    &  0.22$\pm$0.02 \\ 
J0804$+$3012 &0.39$\pm$0.03& 130.1&0.72$\pm$0.06  &1.19069    &  86$\pm$47   & 0.002$\pm$0.001    &  0.03$\pm$0.02 \\ 
             &             &      &               &1.19109    &  43$\pm$12   & 0.004$\pm$0.002    &  0.03$\pm$0.02 \\ 
J0808$+$4950 &0.10$\pm$0.02& 23.82&0.18$\pm$0.04  &1.40732    &  11$\pm$2    & 0.008$\pm$0.001    &  0.02$\pm$0.01 \\ 
J0850$+$5159 &15.3$\pm$0.5 & 61.38&28.1$\pm$0.9   &1.32674    &  49$\pm$16   & 0.117$\pm$0.036    &  1.11$\pm$0.50 \\ 
             &             &      &               &1.32692    &  24$\pm$4    & 0.347$\pm$0.108    &  1.61$\pm$0.57 \\ 
J0852$+$3435 &6.91$\pm$0.34& 76.13&12.7$\pm$0.6   &1.30919    &  23$\pm$7    & 0.078$\pm$0.025    &  0.35$\pm$0.15 \\ 
             &             &      &               &1.30945    &  63$\pm$12   & 0.079$\pm$0.009    &  0.96$\pm$0.21 \\ 
J1058$+$4939 &0.41$\pm$0.04& 32.85&0.75$\pm$0.07  &1.21168    &  26$\pm$9    & 0.006$\pm$0.001    &  0.03$\pm$0.01 \\ 
             &             &      &               &1.21181    &  15$\pm$2    & 0.018$\pm$0.001    &  0.05$\pm$0.01 \\ 
J1623$+$0718 &0.91$\pm$0.10& 34.57&1.67$\pm$0.18  &1.33567    &  11$\pm$2    & 0.042$\pm$0.005    &  0.09$\pm$0.02 \\ 
             &             &      &               &1.33585    &  17$\pm$5    & 0.023$\pm$0.005    &  0.08$\pm$0.03 \\ 
J2340$-$0053 &2.50$\pm$0.24& 7.76 &4.59$\pm$0.44  &1.36087    &  2.5$\pm$0.5 & 0.662$\pm$0.207    &  0.32$\pm$0.12 \\ 
             &             &      &               &1.36090    &  5.8$\pm$2.8 & 0.183$\pm$0.042    &  0.21$\pm$0.11 \\ 
J2358$-$1020 &0.23$\pm$0.04& 12.48&0.42$\pm$0.07  &1.17304    &  6$\pm$1     & 0.034$\pm$0.004    &  0.04$\pm$0.01 \\ 
\hline
\multicolumn{8}{l}{$^a$ in units of 10$^{18}$ cm$^{-2}$; $^b$ FWHM in km s$^{-1}$; $^c$ peak optical depth; $^d$ in units of 10$^{19}$ cm$^{-2}$}\\
\end{tabular*}
\label{dettab2}
\end{table*}
%
In this Section, we give details on the nine 21-cm absorbers detected in our survey (Fig.~\ref{mg2det}). 
The total integrated 21-cm optical depths and velocity spreads of the absorption ($\Delta V$) 
are listed in Table~\ref{dettab2}.
We compute $\Delta V$ using the apparent optical depth profile (see Fig.~\ref{aop})
and the method described 
by Ledoux et al. (2006). 
It is usual procedure to decompose the 21-cm absorption profile into multiple Gaussian functions. 
For a Gaussian profile $N$(H~{\sc i}) and the peak optical depth ($\tau_{\rm p}$) are related by,
\begin{equation}
N({\rm HI}) = 1.93 \times 10^{18}~\tau_{\rm p}~\frac{T_{\rm s}}{f_{\rm c}}~\Delta v~{\rm cm}^{-2}.
\end{equation}
Here, $\tau_{\rm p}$ , $\Delta v$, and $f_{\rm c}$ are the peak optical depth, the FWHM in \kms of 
the fitted Gaussian, and the covering factor of the absorbing cloud respectively. Individual 
Gaussian fits and fit parameters are presented in Fig.~\ref{mg2det} and Table~\ref{dettab2} 
respectively. 

%

In principle one can constrain the kinetic temperature $T_{\rm K}$ 
using the width of the absorption lines ($T_{\rm K} \le 21.86\times \Delta v^2$\,K; e.g. Heiles \& Troland 2003). 
However, FWHMs of individual Gaussian components fitted to the absorption profiles are generally broad 
and range from 5\,\kms to 100\,\kms~ when our resolution is $\sim$4~km~s$^{-1}$. 
Most probably these line widths are dominated by processes other than simple thermal broadening
and constraints on $T_{\rm K}$ are poor.  

\noindent
{\bf J0108$-$0037:} The 21-cm absorption profile is well fitted with a 
single Gaussian function with $\tau_{\rm p}$=0.070$\pm$0.002 and $\Delta v=16\pm$1 \kms.  
The absorption line width corresponds to an upper limit on the kinetic temperature of 
$T_{\rm K} \le 5600$\,K. Radio source J0108$-$0037 is resolved into two components separated by $\sim$5\,mas 
($\sim$40\,pc at \zabs~=~1.3710) in the VLBA images at 2 and 8\,GHz (Beasley et al. 2002). Of these 
two the strongest component contributes about 70\% of the total flux.  Since it is difficult to ascertain, 
which of these components is the radio core, estimate of $f_{\rm c}$ is highly uncertain.  
From the integrated optical depth given in Table~\ref{mg2obsres}, we derive 
$N$(H~{\sc i})=2.4$\times 10^{20}$~cm$^{-2}$ for $f_{\rm c}$~=~1 and $T_{\rm s}$~=~100\,K (as in cold neutral 
medium (CNM) of the Galaxy (Kulkarni \& Heiles 1988) ).  
This is a lower limit on $N$(H~{\sc i}) as $f_{\rm c}$ is most probably less than 1 and $T_{\rm s}$ could be larger 
than 100 K. Thus $N$(H~{\sc i}) in the 21-cm component is consistent with this system being a DLA. 
The Zn~{\sc ii}+Cr{\sc ii} blend at $\lambda$~=~2062\,\AA, the Si~{\sc ii} and the Mn~{\sc ii} absorption lines 
are clearly detected even in the low dispersion spectrum (see Fig.~\ref{metalfig}). As can be seen from Table~\ref{mg2sample},
this system has $W_{\rm r}$(Mg~{\sc ii}$\lambda$2796)$\le0.5$ {\AA} and was observed in the early stages of the 
survey mainly because of the presence of strong absorption lines from weak metal transitions (see Fig.~\ref{metalfig}). 
For this reason this system will not figure in any analysis we perform using equivalent width limited samples. 
 
\noindent
{\bf J0804$+$3012:} 
Broad and shallow 21-cm absorption detected towards this QSO was confirmed by observations 
at different epochs and using a wider baseband bandwidth of 2\,MHz. Therefore, latter spectra have a 
factor 2 coarser velocity resolution (see Table~\ref{mg2obslog}). In Fig.~\ref{mg2det}, we show the weighted average 
of the spectra obtained after smoothing the spectrum of higher resolution. This system has the broadest velocity 
profile among the 21-cm absorbers in our sample.
The overall absorption profile is fitted with a main component with $\Delta v\sim$43 \kms and a broad wing, $\Delta v\sim$86 \kms.  
Based on the integrated optical depth we derive $N$(H~{\sc i})=7.0$\times$10$^{17}$($T_{\rm s}/f_{\rm c}$)\,cm$^{-2}$.

This source is resolved into three components in the MERLIN images obtained at 1.6 GHz and 5 GHz
with angular resolutions 0.178$^{\prime\prime}\times$0.155$^{\prime\prime}$ and 
0.073$^{\prime\prime}\times$0.037$^{\prime\prime}$ respectively (Kunert et al. 2002). 
The angular extent of the radio emission in these images is $\sim 1.4''$
with 77\% of the flux contributed by the central component. However, at 1.4 GHz these 
three components account for only 50\% of the flux measured in FIRST images. 
The structure of the source is thus highly complex. This could very well explain the large 
velocity spread of the 21-cm absorption. This source is ideal for
VLBA spectroscopy to study the internal structure of the intervening Mg~{\sc ii} system.
Apart from Mg~{\sc ii}, Fe~{\sc ii} and Mg~{\sc i} absorption lines no other
metal absorption lines are detected in the SDSS spectrum (see Fig.~\ref{metalfig}).

\noindent
{\bf J0808$+$4950:}
The background QSO at \zem = 1.432 is classified as highly polarized Blazar/QSO with radio 
components at mas scales showing superluminal motion (see Piner et al. 2007). 
The measured apparent velocity is $v_{\rm a}\sim 15$c with $c$ being the speed of light.
This source is also known to show strong flux variability at high radio frequencies 
(Terasranta et al. 2005). We detect a weak 21-cm absorption towards this quasar with
absorption redshift \zabs = 1.407, only $\sim$3000 \kms away 
from the QSO emission redshift. Absorption was confirmed by repeated observations 
at epochs separated by seven months. We do not note any strong flux variation
between these observations. The spectrum presented in Fig.~\ref{mg2det} 
results from the combination of all data. The absorption profile is well fitted with a single Gaussian 
component with $\tau_{\rm p}$=0.008$\pm$0.001 and $\Delta v$=11$\pm$2\,\kms.
This source is unresolved at 8\,GHz from MERLIN observations part of the
CLASS/JVAS survey with a beam size of 0.227$^{\prime\prime}\times$0.203$^{\prime\prime}$. VLBA observations
at 2 and 8 GHz resolve the source into two components with a separation of
2 and 1 mas respectively (Fey \& Charlot 2000). The strongest component accounts
for 85\% of the flux detected in the VLBA map. However the total flux density
480 mJy measured at 8 GHz in the VLBA observations accounts only for 55\% 
of the flux detected in the MERLIN image. It is not clear whether this 
difference is due to flux variations or to a considerable fraction of the
flux being emitted in a diffuse component that is resolved in the VLBA images.
The H~{\sc i} column density derived from our GMRT observations is, 
$N$(H~{\sc i})=2.2$\times 10^{19}(T_{\rm s}/100)(0.85/f_{\rm c})$\,cm$^{-2}$.     
Only Mg~{\sc ii}, Fe~{\sc ii} and Mg~{\sc i} absorptions are detected in the
SDSS spectrum (See Fig.~\ref{metalfig}). 

\noindent
{\bf J0850$+$5159:}
Radio source J0850+5159 is compact at mas scales in the VLBA images at 5 and 15\,GHz (Taylor et al. 2005) 
implying $f_{\rm c}\sim$1.0. 
Strong ($\tau_{\rm p}$=0.50) 21-cm absorption is detected towards this QSO which is extremely red at optical and 
NIR wavelengths (u-K$\sim$4.8\,mag).  
The total 21-cm H~{\sc i} column density, with $f_{\rm c}=1$, is 
$N$(H~{\sc i})~=~(2.81$\pm$0.09)$\times 10^{19} T_{\rm s}$\,cm$^{-2}$.
Numerous absorption lines are detected in the SDSS spectrum in addition to 
Mg~{\sc ii}, Fe~{\sc ii} and Mg~{\sc i} (see Fig.~\ref{metalfig}).
A 2175\,{\AA} feature is also present at 2.7$\sigma$ significance level at the redshift of 
the intervening \mgii system (Srianand et al. 2008b).  
An independent estimate of $N$(H~{\sc i}) was derived by fitting a LMC2 extinction curve to the 
QSO spectrum, $N$(H~{\sc i})~=~5.73$\pm$1.10$\times$10$^{21}$~cm$^{-2}$.
Therefore we conclude that $T_{\rm s}\sim 204 \pm 40$\,K~ (see Srianand et al. 2008b for more details). 
 
\noindent
{\bf J0852$+$3435:} 
Broad and shallow 21-cm absorption, confirmed by observations at three different epochs, is detected 
towards this red QSO (u-K$\sim$5.6\,mag). Details about the 2175\,{\AA} feature detected 
at $>$ 5$\sigma$ significance level at the redshift of the Mg~{\sc ii} absorption system
are presented in Srianand et al. (2008b). 
Numerous absorption lines are detected in the SDSS spectrum in addition to
Mg~{\sc ii}, Fe~{\sc ii} and Mg~{\sc i} (see Fig.~\ref{metalfig}).  
The background radio source is compact in CLASS/JVAS and has flat radio spectrum (Jackson et al. 2007).
Total 21-cm H{\sc i} column density with $f_{\rm c}=1$ is, $N$(H~{\sc i})=(1.27$\pm$0.06)$\times 10^{19} T_{\rm s}$\,cm$^{-2}$.
An independent estimate of $N$(H~{\sc i}) was derived by fitting the LMC2 extinction curve to the QSO spectrum 
$N$(H~{\sc i})~=~6.97$\pm$1.30$\times$10$^{21}$~cm$^{-2}$.
Therefore we conclude that $T_{\rm s}\sim 550\pm 106$\,K (Srianand et al. 2008b).\footnote{Note errors in $T_{\rm s}$ measurements for this system 
and the system at \zabs = 1.326 towards J0850+5159 are slightly reduced compared to that given in Srianand et al. (2008b) because
of different procedures (we use here the errors obtained from the apparent optical depth profile technique).}

\noindent
{\bf J1058$+$4939:} 
The 21-cm absorption line detected towards this QSO is fitted well with two Gaussian components.
The radio source is unresolved in the FIRST and GMRT images. The quasar is known to show
variability at 151 MHz. Riley \& Warner (1994) found that roughly 60\% of the total flux at 
5 and 8\,GHz comes from a region smaller than $\le$ 0.1$^{\prime\prime}$ implying $f_{\rm c}$=0.6.  
VLBA or MERLIN images are not available for this source. Integrated 21-cm optical depth yields
$N$(H~{\sc i})~=~1.25$\times 10^{20} (T_{\rm s}/100)(0.60/f_{\rm c})$\,cm$^{-2}$. 
The SNR of the SDSS spectrum is poor and only a few absorption lines
are detected (see Fig.~\ref{metalfig}).

\noindent
{\bf J1623$+$0718:} 
21-cm absorption is detected in two well detached components.
The background radio source is unresolved in GMRT and FIRST images. 
The  source is compact even in CLASS/JVAS and has a flat radio spectrum 
(Jackson et al. 2007). Thus $f_{\rm c}$ could be close to 1 at $\sim$100 mas scales.
The integrated 21-cm optical depth yields $N$(H{\sc i})~=~1.7$\times 10^{18} T_{\rm s}$\,cm$^{-2}$.
The Zn~{\sc ii}+Cr{\sc ii} blend  at $\lambda$=2062\,\AA, the Si~{\sc ii} and the Mn~{\sc ii} 
absorption lines are clearly detected 
even in the SDSS low dispersion spectrum (see Fig.~\ref{metalfig}).
 
\noindent
{\bf J2340$-$0053:} 
Radio source J2340$-$0053 is compact at mas scales in VLBA observations at 2 and 8\,GHz 
(Kovalev et al. 2007). 
The 21-cm absorption is narrow and the spectrum presented in Fig.~\ref{mg2det}
was obtained with a resolution of 1.0 \kms~ per channel.
The main component has a FWHM of 2.5$\pm$0.5 \kms~
which corresponds to an upper limit of 136~K for the kinetic temperature. 
Clearly the gas  responsible for this system is cold. 
It is most likely that $f_{\rm c}\sim 1$ for this system. Under the assumption
that $f_{\rm c} =1$ and $T<200$~K, we derive $N$(H~{\sc i})~$\le$ 9$\times 10^{19}$ 
cm$^{-2}$. This means the component responsible for this 21-cm 
absorption is most probably a sub-DLA. An optical spectrum of this QSO
was analysed by Khare et al. (2004). 
\begin{figure}
\centerline{\hbox{
\psfig{figure=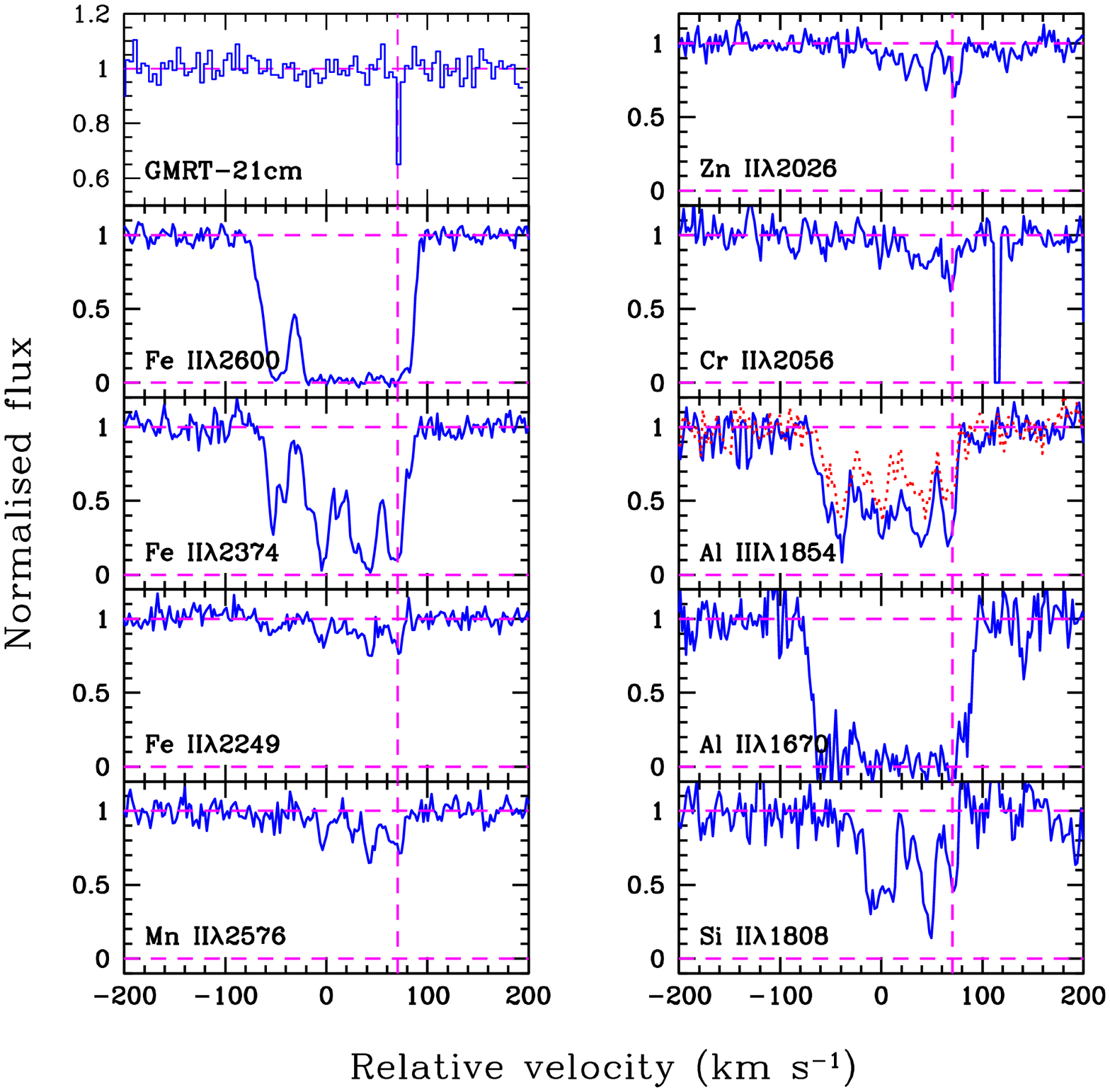,height=8.0cm,width=8.0cm,angle=0}
}}
\caption[]{Absorption lines in the \zabs = 1.3609 system
towards J2340$-$0053 observed at high spectral resolution 
(Prochaska et al. 2007). The dotted line spectrum corresponds to
Al~{\sc iii}$\lambda 1862$. 
Dashed vertical lines show the location of the 21-cm absorption line.}
\label{j2340}
\end{figure}
In Fig.~\ref{j2340} we show absorption lines associated to this 
systems on a velocity scale.
The corresponding high spectral resolution KECK spectrum was obtained by 
Prochaska et al. (2007). The vertical
dashed lines mark the location of the 21-cm absorption. It is clear
that the 21-cm absorption is not originating from the component that
dominates the equivalent width of strong metal lines.
Weak metal lines such as Fe~{\sc ii}$\lambda$2249 
suggest that $\sim 30$\% of the metals are in this component.
The compactness of the background quasar and the fact that the metal absorption
component associated with 21-cm is well detached from the central blend
makes it an interesting system for measuring the time evolution of
fundamental constants. However, the publicly available KECK data is not
sufficient for this purpose. 

\noindent
{\bf J2358$-$1020:} 
The background radio source is unresolved in high frequency VLBA observations (Fomalont
et al. 2000; Fey \& Charlot, 2000) and $f_{\rm c}$ could be close to unity.
The absorption profile is well fitted with a single Gaussian component 
whose FWHM implies a kinetic temperature $\le$ 800 K. This gives
log[$N$(H~{\sc i})]$\le20.5$ for $f_{\rm c}$ = 1. The 
Zn~{\sc ii}+Cr~{\sc ii}$\lambda$2062\AA~ blend is clearly detected in the low dispersion SDSS
spectrum (Fig.~\ref{metalfig}).

\section{Detectability of 21-cm absorption}
\begin{figure*}
\centerline{
\vbox{
\hbox{
\psfig{figure=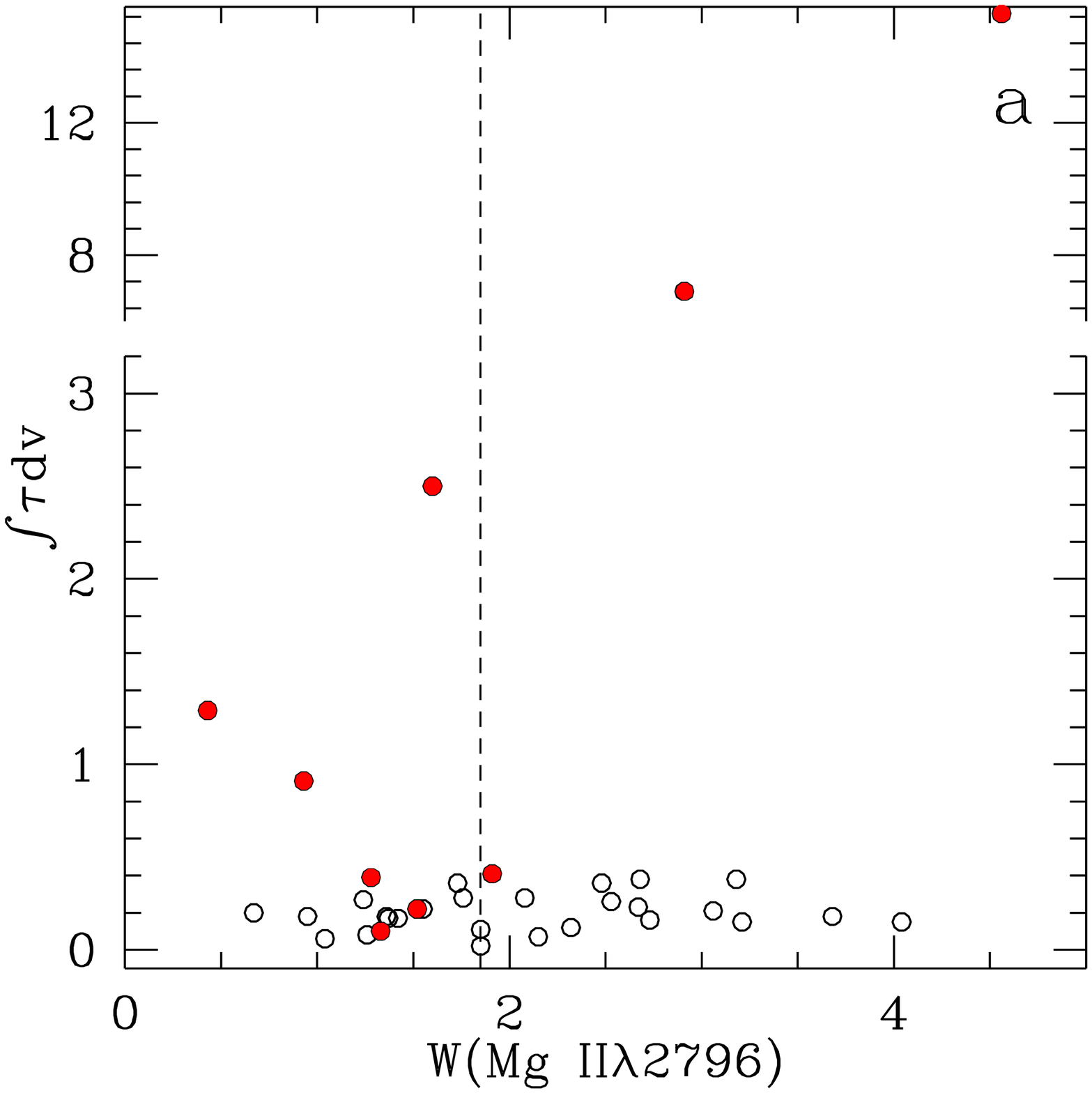,height=4.3cm,width=4.3cm,angle=0}
\psfig{figure=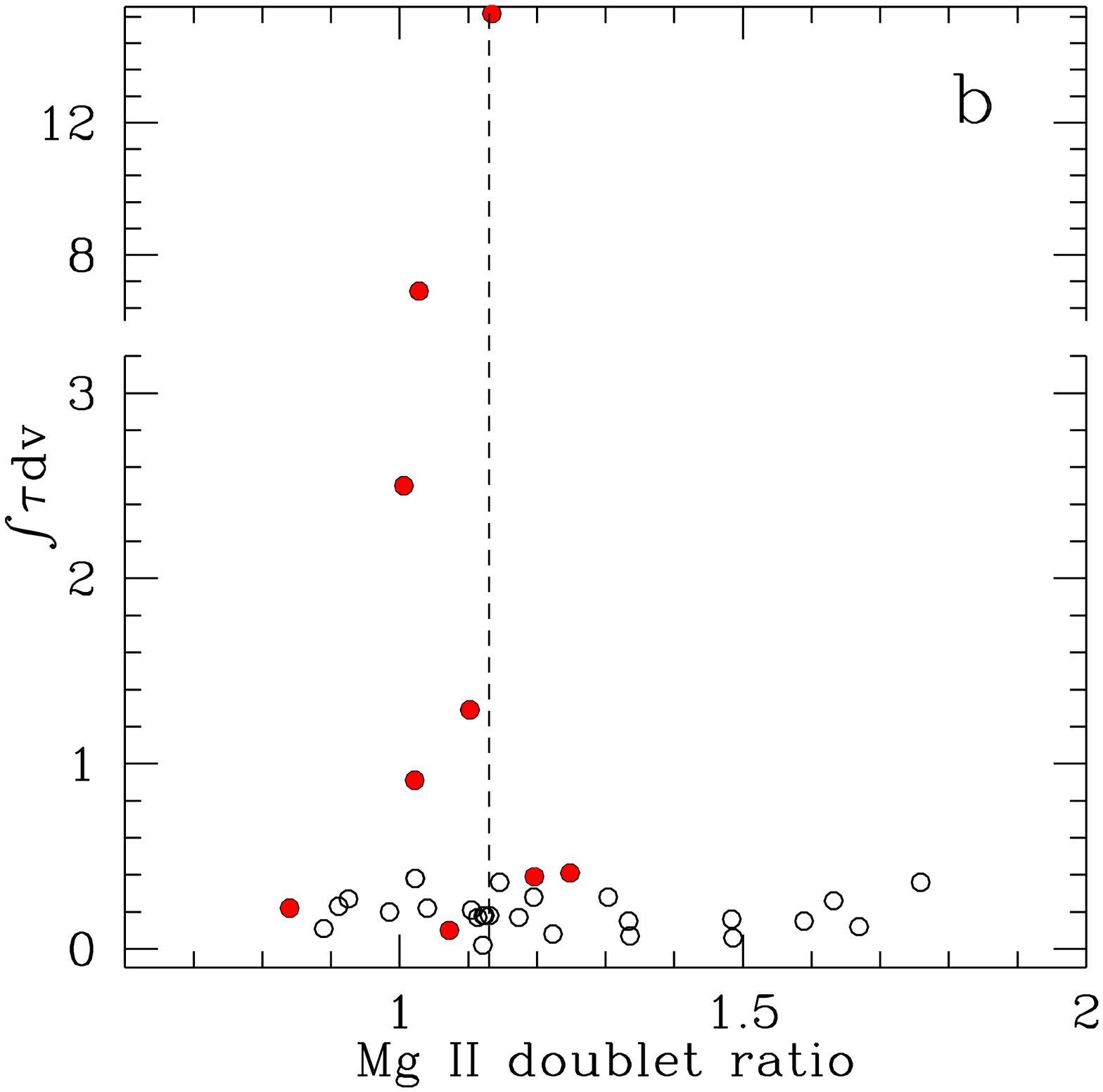,height=4.3cm,width=4.3cm,angle=0}
\psfig{figure=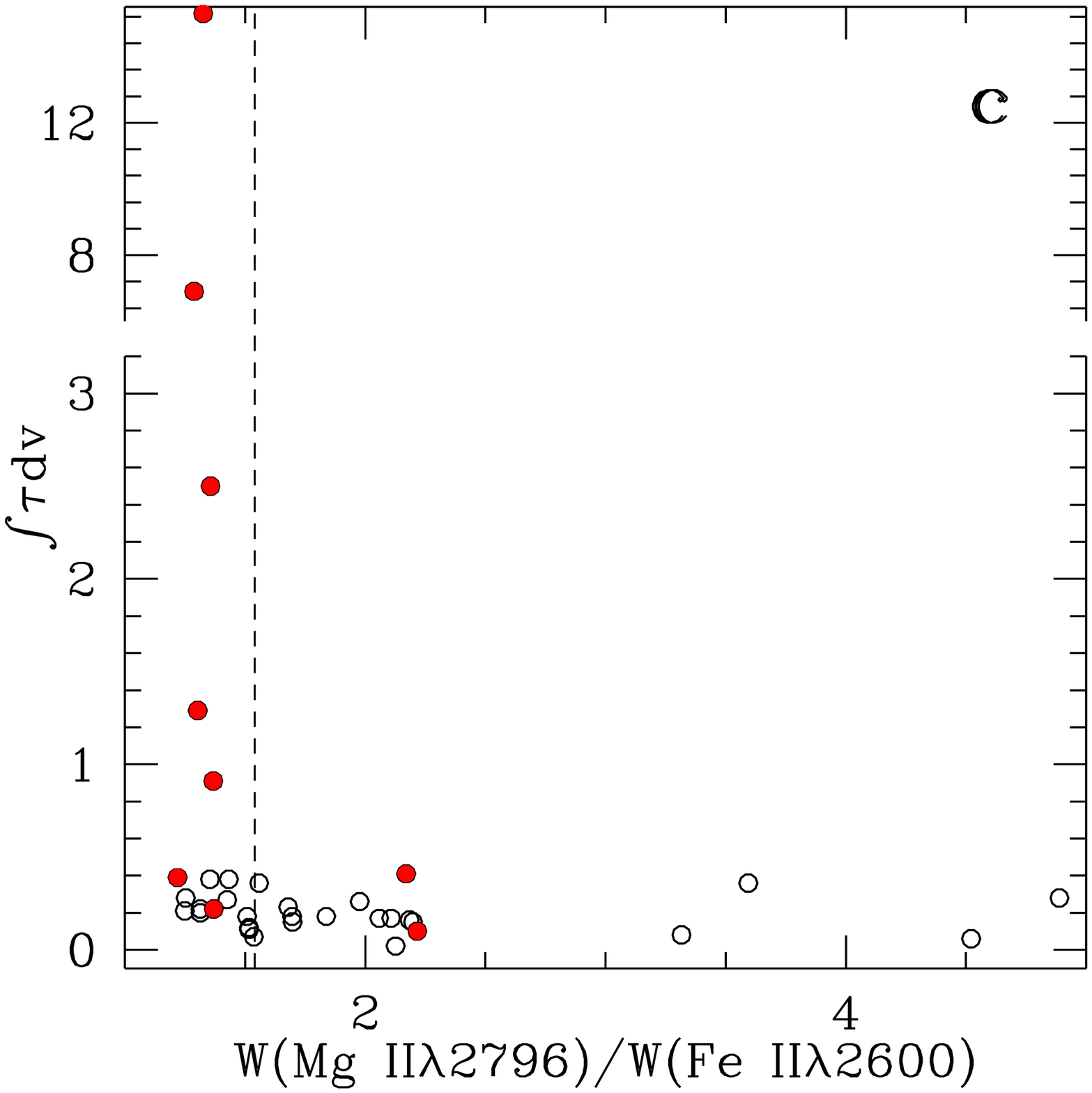,height=4.3cm,width=4.3cm,angle=0}
\psfig{figure=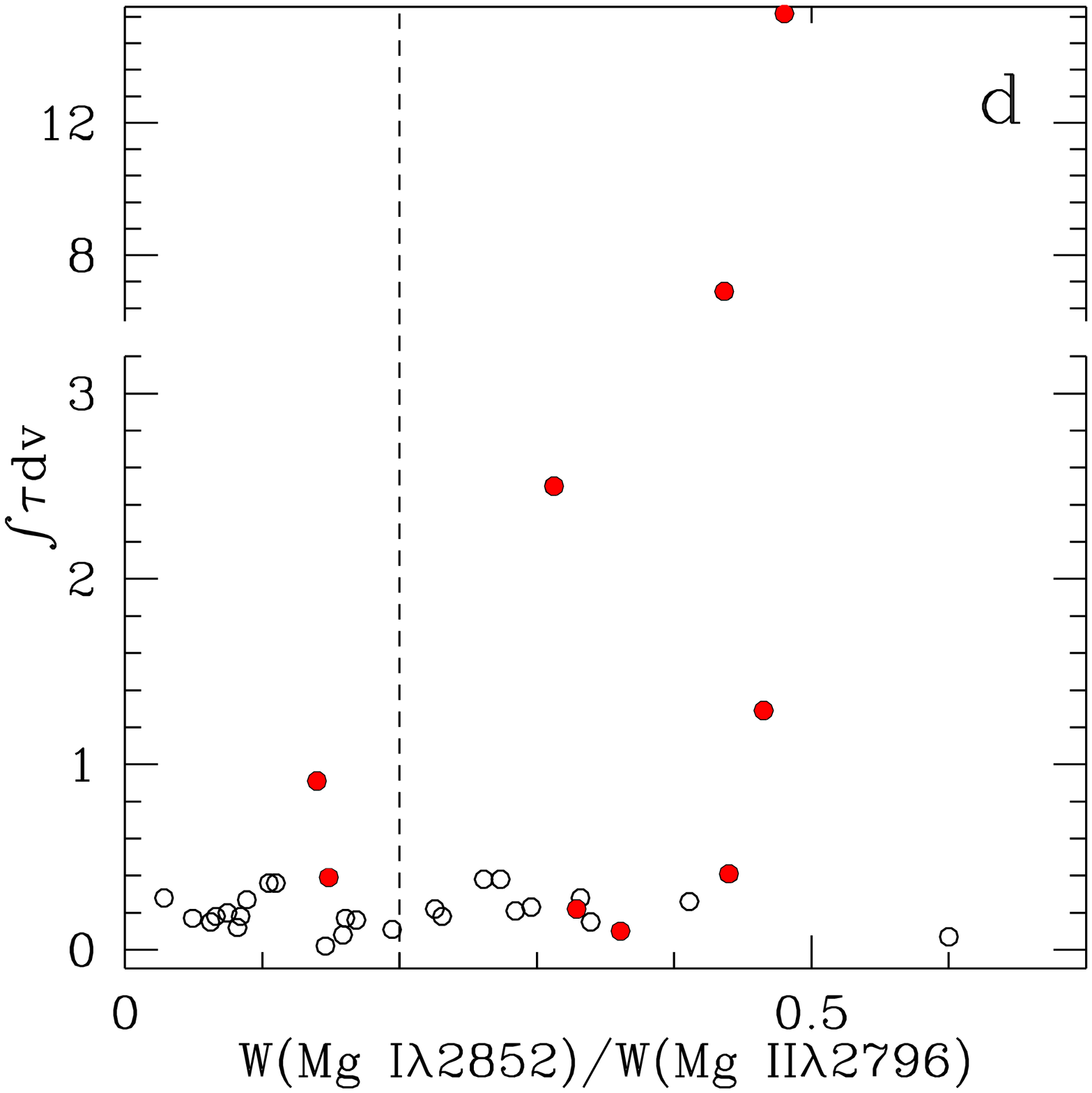,height=4.3cm,width=4.3cm,angle=0}
}
\hbox{
\psfig{figure=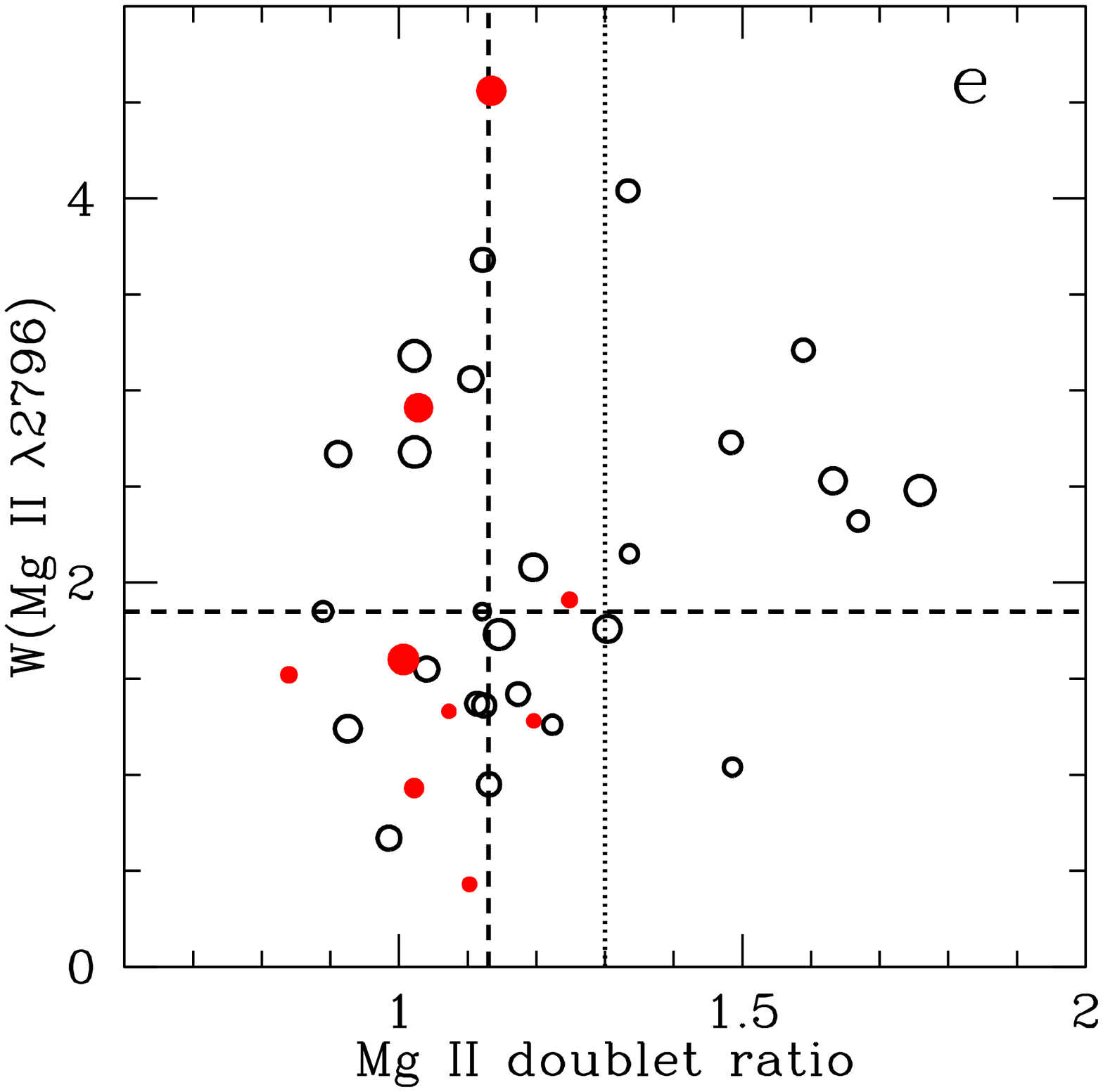,height=4.3cm,width=4.3cm,angle=0}
\psfig{figure=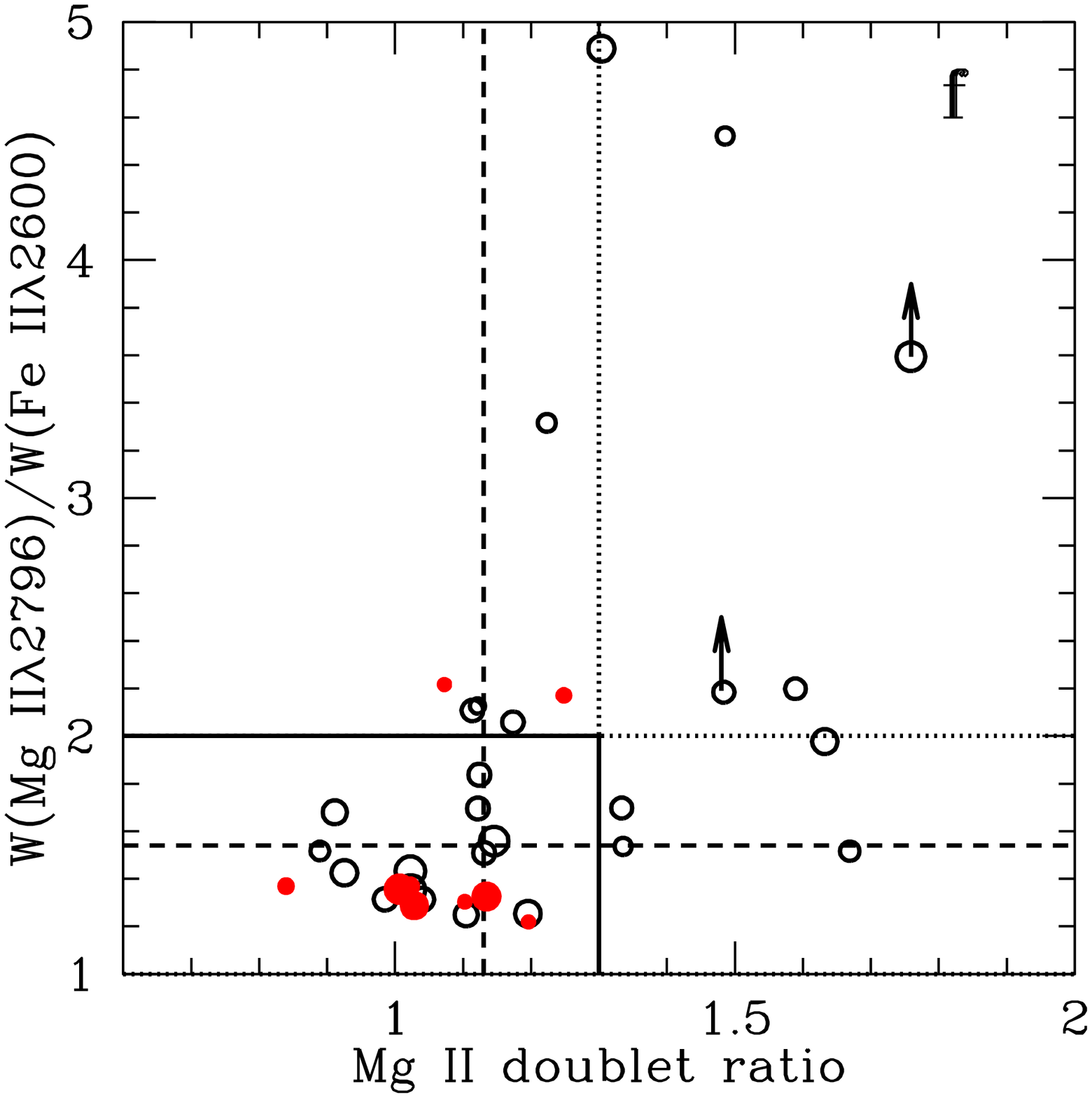,height=4.3cm,width=4.3cm,angle=0}
\psfig{figure=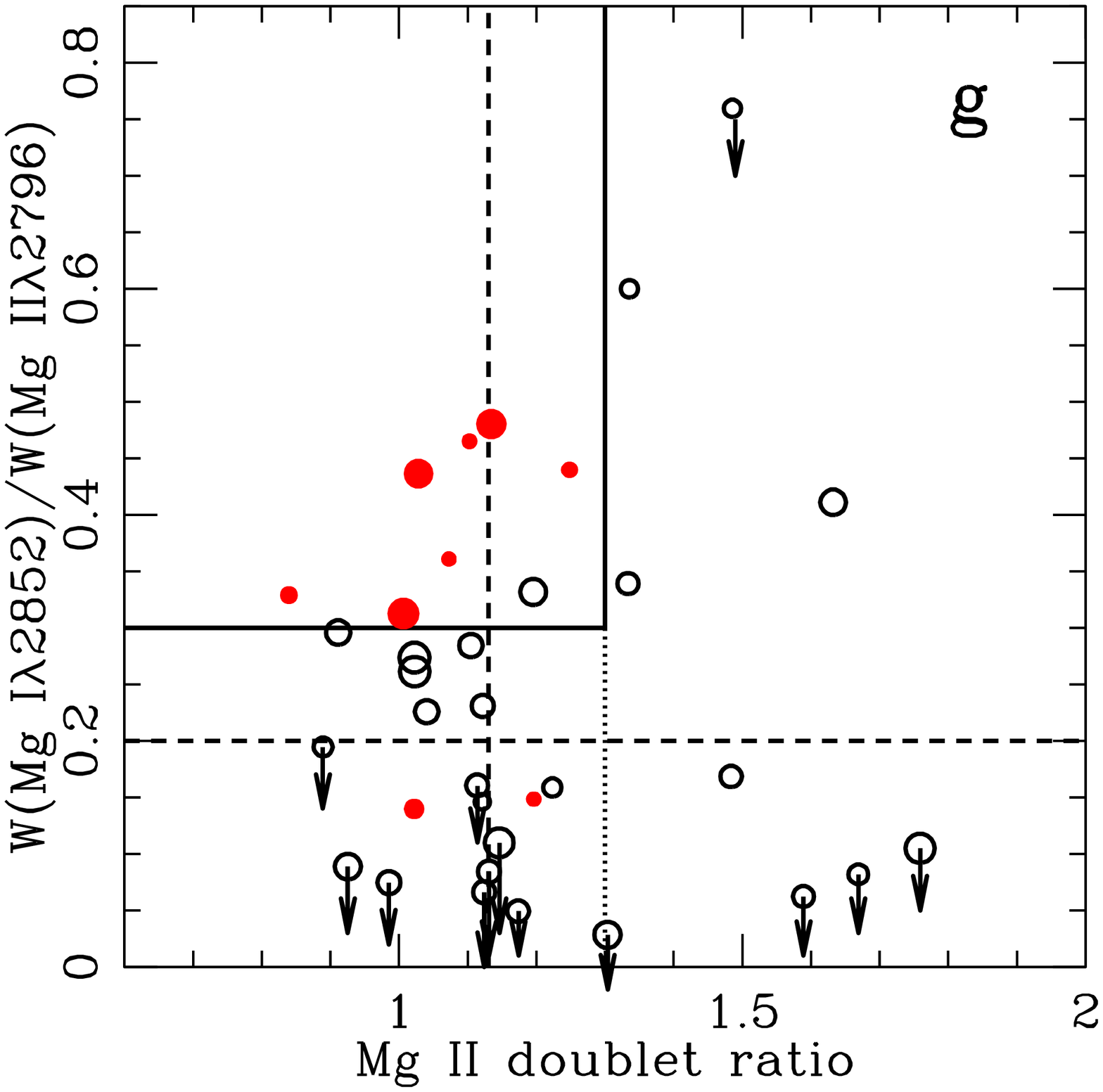,height=4.3cm,width=4.3cm,angle=0}
\psfig{figure=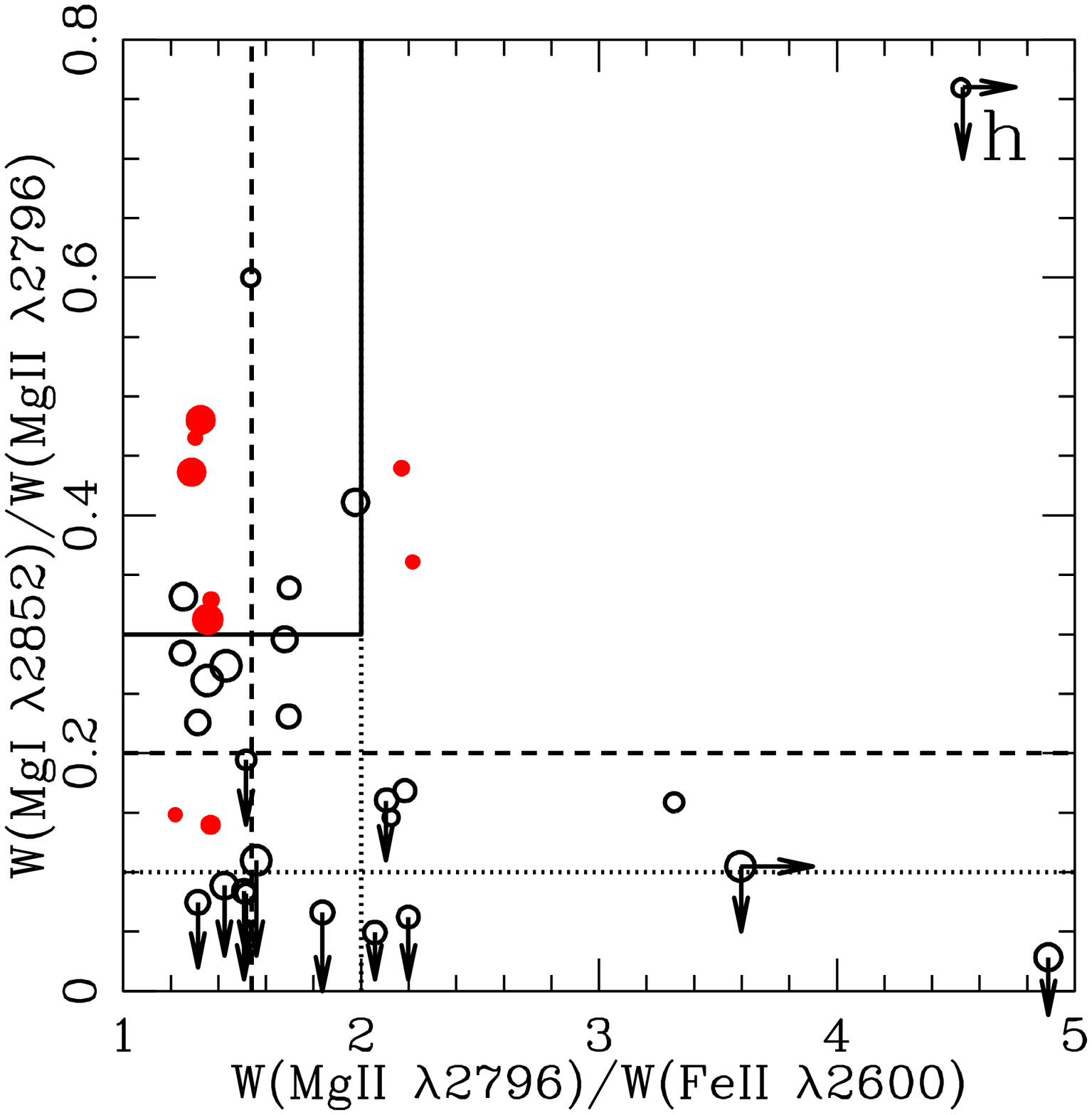,height=4.3cm,width=4.3cm,angle=0}
}
}
}
\caption[]{{\it Top panels:}
The integrated 21-cm optical depth is plotted against various
parameters of the UV absorption lines derived from the SDSS spectra. 
In the case of detection (filled circle) the optical depth is obtained
by integrating over the observed absorption profile. Open circles are
for 21-cm non-detections in which case limits
are obtained by integrating the optical depth over a Gaussian
component with peak optical depth corresponding to the 3$\sigma$ rms limit in 
the continuum and width 10 km~s$^{-1}$.
{\it Bottom panels:} Different parameters of UV absorption lines are
plotted against one another. Symbols are as in upper panels.
The size of the symbols are scaled by the 21-cm optical 
depth limits (per 10\,km\,s$^{-1}$) reached. The vertical dashed line in all the panels gives the
median value. Other lines indicate the limiting values and/or
allowed ranges as discussed in the text.
}
\label{mg2hist}
\end{figure*}

In this Section, we investigate the relationship between
the detectability of 21-cm absorption and properties of UV absorption lines
from the Mg~{\sc ii} systems. 
This may help in particular to preselect more efficiently candidates 
for future 21-cm surveys.

\subsection{Rest equivalent width of Mg II absorption}

First we concentrate on $W_{\rm r}$(Mg~{\sc ii}$\lambda$2796), the only
quantity that has been used to define our present sample.
In panel (a) of Fig.~\ref{mg2hist} we plot 
$W_{\rm r}$(Mg~{\sc ii}$\lambda$2796) against the integrated 21-cm optical
depth for all the systems in our GMRT sample. 
Detections (resp. non detections) are marked by filled (resp. open) circles.
Even though the system with strongest 21-cm absorption in our sample
has the largest equivalent width, we do not find evidence for 
a correlation between the two quantities.
Indeed, the median $W_{\rm r}$(Mg~{\sc ii}$\lambda$2796) is 1.85{\AA} for the systems
observed with GMRT and there are only three (out of 9) 21-cm detections from systems
having  $W_{\rm r}$(Mg~{\sc ii}$\lambda$2796)$\ge$1.85{\AA}. 
As we have reached similar limiting 21-cm optical depth over the whole 
range of $W_{\rm r}$(Mg~{\sc ii}$\lambda$2796) sampled by our survey, this 
can not be due to any artifact introduced by variation of the optical depth limits
through the sample. 

For comparison with our sample, we summarize in Table~\ref{lanesamp} the properties of 
systems with $W_{\rm r}$(Mg~{\sc ii}$\lambda$2796)$\ge$ 1 {\AA} in the low-$z$ sample 
of Lane (2000). The median $W_{\rm r}$(Mg~{\sc ii}$\lambda$2796)  and redshift in this
subsample is {1.75} {\AA} and \zabs=0.7.
There are 2 and 5 detections respectively in systems with $W_{\rm r}$(Mg~{\sc ii}$\lambda$2796) 
below and above the median value. It seems therefore that the probability to detect 21-cm 
absorption in strong Mg~{\sc ii} systems is higher at lower redshift.
We further investigate this issue in the next Section.
%

\subsection{Mg~{\sc ii} Doublet ratio (DR)}

In panel (b) of Fig.~\ref{mg2hist} we plot  the integrated 21-cm
optical depth vs. the Mg~{\sc ii} doublet ratio (DR). It is clear that 21-cm detected 
systems are confined  to regions with DR$<$1.3. For the systems with 
$W_{\rm r}$(Mg~{\sc ii}$\lambda$2796)$\ge$ 1{\AA} and DR$<$1.3 we detect 21-cm 
absorption in roughly {33$^{+17}_{-12}$\%  (8/24)} of the systems to be compared
to the overall success rate in the sample, {24$^{+12}_{-8}$\% (8/33)}.
The median value of DR in our sample is 1.12 and only two systems with DR greater than this are 
detected in 21-cm absorption. 

As pointed out by Petitjean \& Bergeron (1990)
the number of Voigt profile components found in high resolution 
spectra correlates well with the total equivalent width of blended lines
(see Fig. 3 of Ellison 2006). Thus equivalent widths measured in SDSS data 
reflect primarily the number of individual components and/or the velocity 
spread between them and 
not directly the column density (see P06).
On the contrary, a doublet ratio close to 1, even in a low dispersion
spectrum, means high level of saturation and therefore high column density.
Thus systems with high $W_{\rm r}$(Mg~{\sc ii}$\lambda$2796) and DR$\sim$1 provide higher probability
of a line of sight hitting high Mg~{\sc ii} column density clouds.

Interestingly, in the GMRT sample, there are 9 systems with DR$>$1.3 and 7 of these systems 
have $W_{\rm r}$(Mg~{\sc ii}$\lambda$2796) $>$ 1.85~\AA~ (see panel e of Fig.~\ref{mg2hist}). 
The lower detection rate of 21-cm absorption in systems
with  $W_{\rm r}$(Mg~{\sc ii}$\lambda$2796)$>$ 1.85 in the GMRT sample could be due to high values
of DR in most of these systems. It is most likely that the high value
of $W_{\rm r}$(Mg~{\sc ii}) in these systems is primarily due to the velocity field and
not to a large Mg~{\sc ii} column density. 

From Fig.~3.3 of Lane (2000) we find that all the low-$z$ 21-cm detections have DR$\le$1.4. 
Thus our finding of high detection rate of 21-cm absorption in systems with low DR is consistent with what is seen for 
low-$z$ 21-cm absorbers. 
The difference between high and low redshift sample noted in the previous Section is therefore probably
related to a redshift evolution in the overall structure of Mg~{\sc ii} systems.

\subsection {The $W_{\rm r}$(Mg~{\sc ii}$\lambda$2796)/$W_{\rm r}$(Fe~{\sc ii}$\lambda2600$) ratio}

RTN06 showed that DLAs have 1.0$\le$ 
$W_{\rm r}$(Mg~{\sc ii}$\lambda$2796)/$W_{\rm r}$(Fe~{\sc ii}$\lambda2600$)$\le$ 2.
We shall call this ratio $R_1$.
In their Mg~{\sc ii} sample there are {19} systems in the redshift range 
$1.10\le z\le 1.45$, with $W_{\rm r}$(Mg~{\sc ii}$\lambda$2796)$\ge 1${\AA}
and  $R_1$ in the above mentioned range. 
Seven of them (i.e 37$^{+20}_{-14}$\%) have log $N$(H~{\sc i})$\ge$20.3 and 
are DLAs as per the conventional definition. However, 18 of these  
19 systems have log $N$(H~{\sc i})$\ge$19.6.

From Panel (c) of Fig~\ref{mg2hist} it is clear that our selection
of systems based on $W_{\rm r}$(Mg~{\sc ii}$\lambda$2796) alone has ensured
that a large fraction of systems (24/33) in our sample have $R_1$ in the
above mentioned range. 
In addition, the detection rate of 21-cm absorption in our sample is high when 
$R_1$ is low. 
The median value of $R_1$ in our sample is 1.54. About
41$^{+22}_{-15}$\% (resp. 11$^{+15}_{-7}$\%) of the systems with 
$R_1<1.54$ (resp. $>$1.54) show 21-cm absorption.
In the $R_1$ range covered by DLAs in RTN06 we find that there are 
24 Mg~{\sc ii} systems in our sample with 7 of them (i.e 29$^{+16}_{-11}$\%) 
showing detectable 21-cm absorption. The other two detections 
have $R_1$ consistent with 2 within measurement uncertainties. 
For comparison, in the low-$z$ sample of Lane (2000) there are 11 systems 
with $1\le R_1\le 2$ and 6 of them show 21-cm absorption.

In panel (f) of Fig.~\ref{mg2hist} we plot $R_1$ vs DR. 
Out of the 20 systems with both DR$<$1.3 and $R_1<2$, 7 systems show 21-cm absorption
(i.e 35$^{+19}_{-13}$\%) systems. This is very close to the detection
rate we obtained with the condition DR$<$1.3 only. Reducing the upper limit of $R_1$ to 1.6 
produces an increase of the detection rate to 41$^{+22}_{-15}$\%. 

Finally we notice that roughly half the systems with $W_{\rm r}$(Mg~{\sc ii}$\lambda$2796)$>$ 1.85{\AA} 
also have $R_1>$ 2. This again suggests that in this redshift range large 
$W_{\rm r}$(Mg~{\sc ii}$\lambda$2796) does not guarantee high column density.

\subsection {The $W_{\rm r}$(Mg~{\sc i}$\lambda$2852)/$W_{\rm r}$(Mg~{\sc ii}$\lambda2796$) ratio}

From the data listed in Table~1 of RTN06 we find that the DLA
population with $W_{\rm r}$(Mg~{\sc ii}$\lambda$2796)$\ge$ 1{\AA} has 
$W_{\rm r}$(Mg~{\sc i}$\lambda$2852)/$W_{\rm r}$(Mg~{\sc ii}$\lambda$2798)
(called $R_2$ from now on) in the range 0.01 to 0.6. 
Conversely, 
$\sim$30$^{+7}_{-6}$\% of systems with 0.01~$<$~$R_2$~$<$~0.6 are DLAs.
The fraction increases to $\sim$52$^{+19}_{-14}$\% for $R_2>0.3$.
In addition, 92$^{+8}_{-19}$\% of the systems with 
$R_2>0.3$ have log $N$(H~{\sc i})$>$19.60.
From the sample of Lane (2000) we find that 35\% (resp. 40\%) of the systems 
with $W_{\rm r}$(Mg~{\sc ii}$\lambda$2796)$>$1{\AA} and $R_2>0.1$  (resp. $R_2>0.3$) 
show detectable 21-cm absorption.
Note however that the Mg~{\sc i}$\lambda$2852 equivalent  width is not 
always available. 
 
It is clear from panel (d) of Fig.~\ref{mg2hist} that all our 21-cm detections, 
in consistency with the results from low-redshift studies, are
associated with $R_2>0.1$. 
We find that the detection rate of 21-cm absorption is $\sim$36$^{+16}_{-12}$\% when $R_2>0.1$. 
About {44$^{+24}_{-16}$\%} of the systems with $R_2>$0.2 (the median value in our sample)
show 21-cm absorption. If we restrict ourself to  $R_2>0.3$ the detection rate in our
sample increases to 70$^{+30}_{-26}$\%. 
It is also apparent from panel (g) of Fig~\ref{mg2hist} that a joint constraint, DR$<1.3$ and $R_2>0.3$
(solid box in this panel), gives a 88$^{+12}_{-31}$\% detection rate (7 out of 8) for 21-cm
absorption. Thus it appears that using joint constraints on DR and $R_2$ can yield 
a very high detection rate of 21-cm absorption. This is confirmed by the low-$z$ data
of Lane(2000) summarized in Table~\ref{lanesamp}. There are only 5 systems (excluding poor 
upper limits on Mg~{\sc i})  with $R_2>0.3$ and 3 of them show detectable 21-cm absorption. 
One of these non-detections (\zabs = 0.9004 towards 1629+120) has DR = 1.6 and the other system (\zabs = 0.7713 
towards 1556$-$245) has $R_1>5$.  

From panel (h) of Fig.~\ref{mg2hist}, it can be seen that joint constraints on $R_1$ and $R_2$ does not 
produce higher detection rate compared to that obtained from the joint constraints on DR and $R_1$. 


\subsection{Milliarcsecond scale radio structure}
We address here possible effects introduced by issues related
to covering factor of the background source that may play an important role on the
detectability of 21-cm absorption
Although VLBI spectroscopy at 467\,MHz of the \zabs = 2.04 DLA towards PKS\,0458-020 suggests that the absorbing 
gas covers a large fraction of the extended ($\ge$17\,kpc) radio components (Briggs et al. 1989),
time-variability of 21-cm optical depth seen in the absorbers towards B0235+164 and B1127$-$145 suggest that absorbing 
gas is patchy at sub-kpc scales (Wolfe, Briggs \& Davis, 1982; Kanekar \& Chengalur, 2001).
In addition, spatially resolved spectroscopy of the three closely spaced ($<$1~arcsec) images of APM 0827+5255 
have revealed a strong variation in Mg~{\sc ii} optical depths over kpc scales 
(Petitjean et al. 2000; Ellison et al. 2004). 

\begin{table}
\caption{Sample of sources with milli-arcsecond scale images}
\begin{tabular}{lccccc}
\hline
\hline
QSO & $f_c$  & LAS  & $W_{\rm r}$(\mgiia) & $ {1\over f_c}\int\tau dv$\\
    & (VLBA) & (mas)& (\AA) & (\kms) \\
 (1) & (2) & (3) & (4) & (5) \\
\hline
J0108$-$0037 &    0.30$-$0.70$^1$ &  5.5   &0.43$\pm$0.05& $ $   1.84$-$4.3\\  
J0240$-$2309 &    0.65$^{2}$      &  10.0  &1.85$\pm$0.002& $\le$ 0.03\\ 
J0259$-$0019 &    1.00$^1$        &$\le$15 &1.76$\pm$0.04& $\le$ 0.28\\  
J0742$+$3944 &    0.58$^3$        &  8.1   &2.48$\pm$0.41& $\le$ 0.62\\  
J0748$+$3006 &    0.41$^3$        &  54.3  &3.68$\pm$0.06& $\le$ 0.44\\  
J0808$+$4950 &    0.85$^3$        &  9.7   &1.33$\pm$0.09& $ $   0.12\\   
J0850$+$5159 &    0.99$^3$        &  3.0   &4.56$\pm$0.12& $ $   15.5\\    
J0953$+$3225 &    0.84$^3$        &  3.2   &1.55$\pm$0.05& $\le$ 0.26\\    
J1126$+$4516 &    0.57$^3$        &  5.6   &1.26$\pm$0.25& $\le$ 0.14\\    
J1208$+$5441 &    0.88$^3$        &  1.2   &1.85$\pm$0.25& $\le$ 0.13\\    
J1508$+$3347 &    {0.74$^3$}        &  11.4  &2.67$\pm$0.08& $\le$ 0.31\\    
J2340$-$0053 &    0.87$^4$        &  13.6  &1.60$\pm$0.04& $ $   2.87\\   
J2358$-$1020 &    0.98$^2$        &  2.2   &1.52$\pm$0.15& $ $   0.24\\   
\hline
\end{tabular}
\begin{flushleft}
$^1$ Beasley et al. 2002 (2 \& 8 GHz); $^2$ Fomalont et al. 2000 (5 GHz); 
$^3$ Helmboldt et al. 2007 (5 GHz); $^4$ Kovalev et al. 2007 (2 \& 8 GHz)
\end{flushleft}
\label{vlbafc}
\end{table}
%
%
\begin{figure}
\centerline{
\vbox{
\psfig{figure=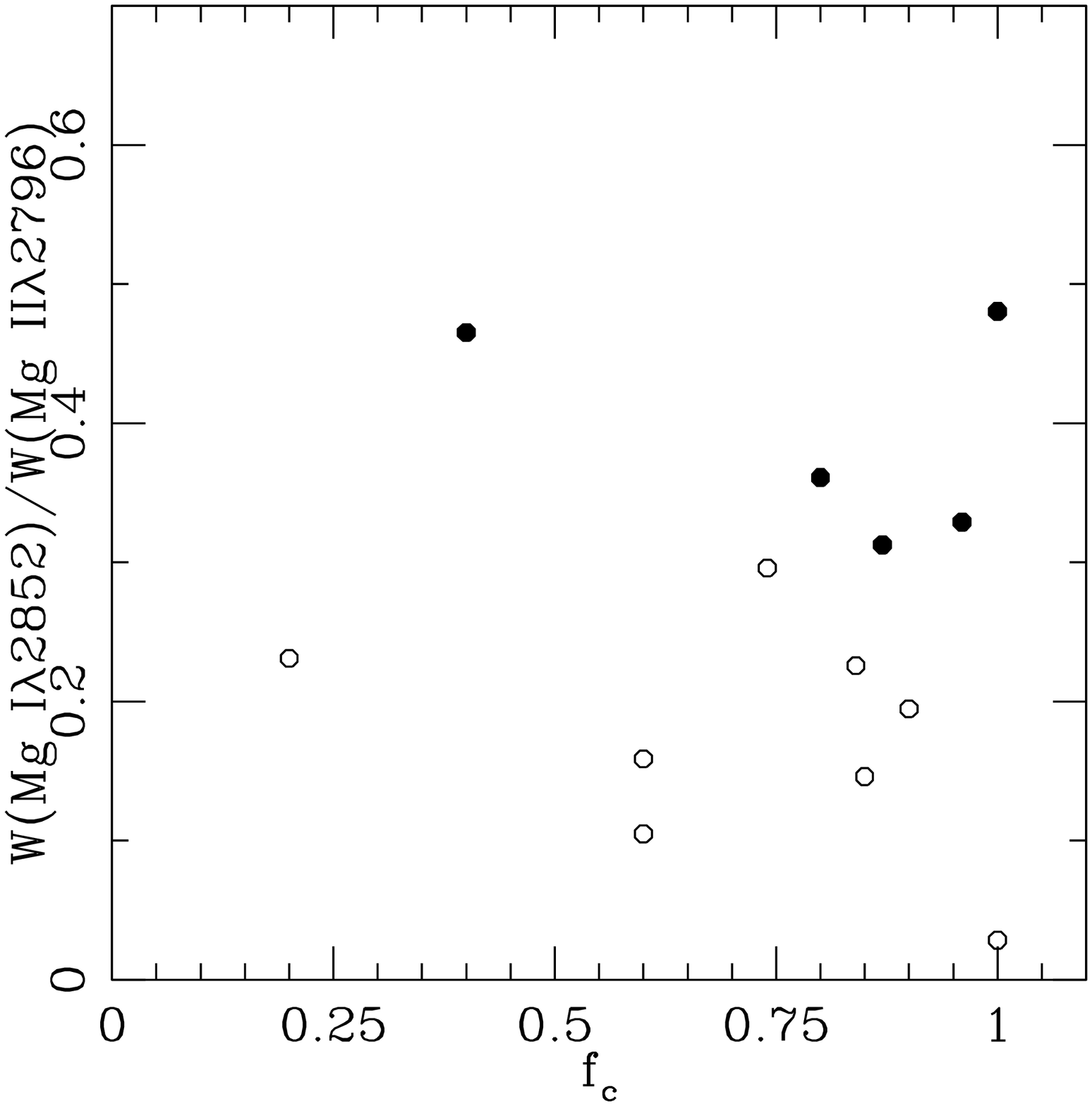,height=8.0cm,width=8.0cm,angle=0}
}
}
\caption[]{The ratio $W_{\rm r}$(Mg~{\sc i}$\lambda$2852)/$W_{\rm r}$(Mg~{\sc ii}$\lambda$2796)
vs. $f_{\rm c}$. The filled and open circles are for systems with and without 
21-cm absorption respectively. 
}
\label{fcvr2}
\end{figure}

\begin{table*}
\caption{Sample of \mgii systems with $W_{\rm r}$(\mgiia)$\ge$1\AA~ from Lane (2000).}
\begin{tabular}{lllccccllr}
\hline
\hline
Source name & \zem    & \zabs   &  $W_{\rm r}$(\mgiia)  & $W_{\rm r}$(\mgiib)  & $W_{\rm r}$(\mgia) & $W_{\rm r}$(\feiia) & dv &   $\tau_{3\sigma}$ & $\int\tau$dv  \\
            &          &        &  (\AA) & (\AA) & (\AA) & (\AA) & (km\,s$^{-1}$) &    & (km\,s$^{-1}$)  \\
  (1)       &    (2)   &   (3)  &  (4)   & (5)   & (6)   & (7)   & (8)            & (9)& (10) \\   
\hline
  0109$+$176 &  2.155  & 0.8392  & 1.75 &   1.20 & $<$0.20  &   1.09   &   9.1             &  0.108  &  $<$1.09 \\ 
  0109$+$200 &  0.746  & 0.5346  & 2.26 &   1.71 &    ....  &   ....   &   7.6             &  0.029  &  $<$0.27 \\ 
  0229$+$341 &  1.240  & 0.7754  & 1.92 &   2.02 & $<$1.13  &   ....   &   8.8             &  0.043  &  $<$0.43 \\ 
  0235$+$164 &  0.940  & 0.5238  & 2.42 &   2.34 &    0.91  &   1.79   &   1.6$^\ddag$     &  0.007  &     13.0 \\ 
  0248$+$430 &  1.316  & 0.3939  & 1.86 &   1.42 &    0.70  &   1.03   &   5.8             &  0.032  &     2.99 \\ 
  0420$-$014 &  0.915  & 0.6330  & 1.02 &   0.86 & $<$0.36  &   ....   &   10.6            &  0.146  &  $<$1.59 \\ 
  0454$+$039 &  1.343  & 0.8597  & 1.53 &   1.40 &    0.37  &   1.11   &   10.0            &  0.010  &  $<$0.11 \\ 
  0735$+$178$^\dag$ &  & 0.4246  & 1.33 &   1.04 &    0.18  &   0.87   &   3.5             &  0.058  &  $<$0.36 \\ 
  0827$+$243 &  0.939  & 0.524   & 2.90 &   2.20 &    ....  &   1.90   &   10.0$^*$        &  0.004  &     0.22 \\ %
  0805$+$046 &  2.876  & 0.9598  & 1.01 &   0.83 & $<$0.60  &   0.24   &   4.9             &  0.095  &  $<$0.71 \\ 
  0957$+$003 &  0.907  & 0.6720  & 1.77 &   1.32 &    ....  &   ....   &   8.3             &  0.105  &  $<$1.01 \\ 
  1127$-$145 &  1.187  & 0.3130  & 2.21 &   1.90 &    1.14  &   1.14   &   5.4             &  0.021  &     3.11 \\ 
  1218$+$339 &  1.519  & 0.7423  & 1.34 &   1.08 & $<$0.70  &   1.00   &   8.6             &  0.020  &  $<$0.20 \\ 
  1229$-$021 &  1.038  & 0.3950  & 2.22 &   1.93 &    0.49  &   1.59   &   1.0$^\ddag$     &  0.007  &     3.00 \\ 
  1327$-$206 &  1.169  & 0.8530  & 2.11 &   1.48 & $<$0.40  &   0.76   &   9.2             &  0.201  &  $<$2.04 \\ 
  1354$+$258 &  2.004  & 0.8585  & 1.00 &   0.86 & $<$0.10  &$<$0.20   &   9.2             &  0.103  &  $<$1.05 \\ 
  1556$-$245 &  2.815  & 0.7713  & 2.07 &   1.91 &    1.07  &$<$0.20   &   8.8             &  0.068  &  $<$0.68 \\ 
  1622$+$238 &  0.927  & 0.6560  & 1.29 &$<$1.69 & $<$0.40  &   1.13   &   10.0$^\ddag$    &  0.004  &     2.20 \\ %
  1629$+$120 &  1.795  & 0.5313  & 1.40 &   1.35 &    0.31  &   0.70   &   0.6$^*$         &  0.004  &     0.49 \\ 
             &    ''   & 0.9004  & 1.06 &   0.67 &    0.44  &   0.63   &   9.4             &  0.080  &  $<$0.82 \\ 
  2212$-$299 &  2.703  & 0.6329  & 1.26 &   1.00 &    0.36  &   ....   &   8.1             &  0.217  &  $<$2.07 \\ 
\hline
\end{tabular}
\begin{flushleft}
Col. 1: source name as given in Lane (2000); col. 2: QSO emission redshift; col. 3: Absorption redshift of \mgii system; 
cols. 4, 5, 6 and 7: Rest equivalent widths of \mgiia, \mgiib, \mgia and \feiia; cols. 8 and 9: spectral resolution 
and 3$\sigma$ uncertainty on the optical depth; 
col. 10: integrated 21-cm optical depth for the detections or $3\sigma$ upper limit to it in case of non-detections. 
The upper limits on $\int\tau$dv have been computed for dv=10\,km\,s$^{-1}$.  \\
$^\dag$ BL Lacertae object a featureless continuum except for the intervening 
absorption line system at \zabs=0.4246 (Carswell et al. 1974).\\
$^\ddag$ 21-cm absorption data for 0235+164, 1229$-$021 and 1622+238 are originally from Wolfe et al. (1978),  
Brown \& Spencer (1979) and Curran et al. (2007) respectively. \\ 
$^*$ 21-cm absorption data for 0827+243 and 1629+120 are from Kanekar et al. (2003). \\ 
\end{flushleft}
\label{lanesamp}
\end{table*}

At present, low frequency (i.e $<$1.4 GHz ) mas scale images are not available for 
any of the sources in our sample. 
Our ongoing VLBA observations at low frequencies will provide better
constraints on $f_{\rm c}$ in the near future.
However, high frequency (i.e $\ge$ 2 GHz) milli-arcsecond scale images are available for 13 sources
out of the 22 sources that are compact in FIRST and our 610\,MHz images.  
We give details of these objects in Table~\ref{vlbafc}. Measured covering fraction ($f_{\rm c}$) and largest angular size
(LAS) are given in columns 2 and 3 respectively. 
Covering fraction has been estimated as the ratio of the core flux density to the total flux density detected in the 
VLBA images.
To achieve as much uniformity as possible, we have used 
the VLBA 5\,GHz (rather than VCS 2 and 8\,GHz) images whenever available. 
From this table it appears that the detectability of 21-cm does not depend on the covering factor in this sub-sample. 
We plot in Fig.~\ref{fcvr2} the ratio $R_2$ against the measured covering factor $f_{\rm c}$ for
systems with (filled circles) and without (open circles) 21-cm detection.
It is apparent that while the distribution of $f_{\rm c}$ is similar between detections and non-detections,
$R_2$ separates the two very nicely. 
%
%

A covering factor less than 1 increases the 
3$\sigma$ optical depth limit by a factor of $1/f_{\rm c}$. In order to quantify the detection rate of 21-cm absorption 
we give in the last column of Table~\ref{vlbafc} the integrated optical depth corrected from partial coverage. 
We come back to this subsample in the next Section while discussing the redshift distribution of 21-cm absorbers.

\section{Redshift distribution of 21-cm absorbers}


\begin{table}
\caption{Redshift distribution of 21-cm absorbers }
\begin{tabular}{lcccccc}
\hline
\hline
Sample & $W_{\rm o}$ &${\cal{T}}_0$ &N&N$_{21}$& $C$ & $n_{21}$\\

(1) & (2) & (3) & (4) & (5) & (6) & (7) \\
\hline
GMRT         & 1.0 &0.10&   9  & 4 &  0.44$^{+0.35 }_{-0.21}$   &  0.114$^{+0.094}_{-0.057}$\\
(\zabs = 1.3)&     &0.30&  26  & 3 &  0.12$^{+0.11 }_{-0.06}$   &  0.030$^{+0.030}_{-0.017}$\\
             &     &0.50&  33  & 4 &  0.12$^{+0.10 }_{-0.06}$   &  0.031$^{+0.026}_{-0.016}$\\
             &     &0.50$^a$&12& 2 &  0.17$^{+0.22 }_{-0.11}$   &  0.044$^{+0.057}_{-0.028}$\\
             & 1.8 &0.30& 14   & 1 &  0.07$^{+0.16 }_{-0.06}$   &  0.006$^{+0.013}_{-0.005}$\\
             &     &0.50& 20   & 2 &  0.10$^{+0.13 }_{-0.07}$   &  0.008$^{+0.010}_{-0.006}$\\
Low-$z$      & 1.0 &0.30&  9   & 5 &  0.56$^{+0.38 }_{-0.24}$   &  0.079$^{+0.055}_{-0.035}$\\
(\zabs = 0.5)&     &0.50&  11  & 4 &  0.36$^{+0.29 }_{-0.17}$   &  0.051$^{+0.042}_{-0.025}$\\
             & 1.8 &0.30&  6   & 4 &  0.67$^{+0.53 }_{-0.32}$   &  0.023$^{+0.020}_{-0.012}$\\
             &     &0.50&  7   & 4 &  0.57$^{+0.45 }_{-0.27}$   &  0.020$^{+0.017}_{-0.011}$\\
\hline
\end{tabular}
\begin{flushleft}
Col. 1: Sample name and mean redshift; col. 2: Equivalent width limit;
col. 3: Threshold of $\int\tau$dv in km\,s$^{-1}$;
col. 4: Number of systems in the sample with $\int\tau_{3\sigma,10}$dv$\le{\cal{T}}_o$; col. 5: Number of 21-cm
absorption detections with integration line depth $\int\tau$dv$\ge {\cal{T}}_o$; col. 6: Fraction of \mgii\,
systems with detectable 21-cm absorption; col. 7: Number density per unit redshift of 21-cm absorbers with
$\int\tau$dv$\ge {\cal{T}}_o$.\\
$^a$ considering only the subsample in Table~\ref{vlbafc}.
\end{flushleft}
\label{tabdndz}
\end{table}

\begin{figure}
\psfig{figure=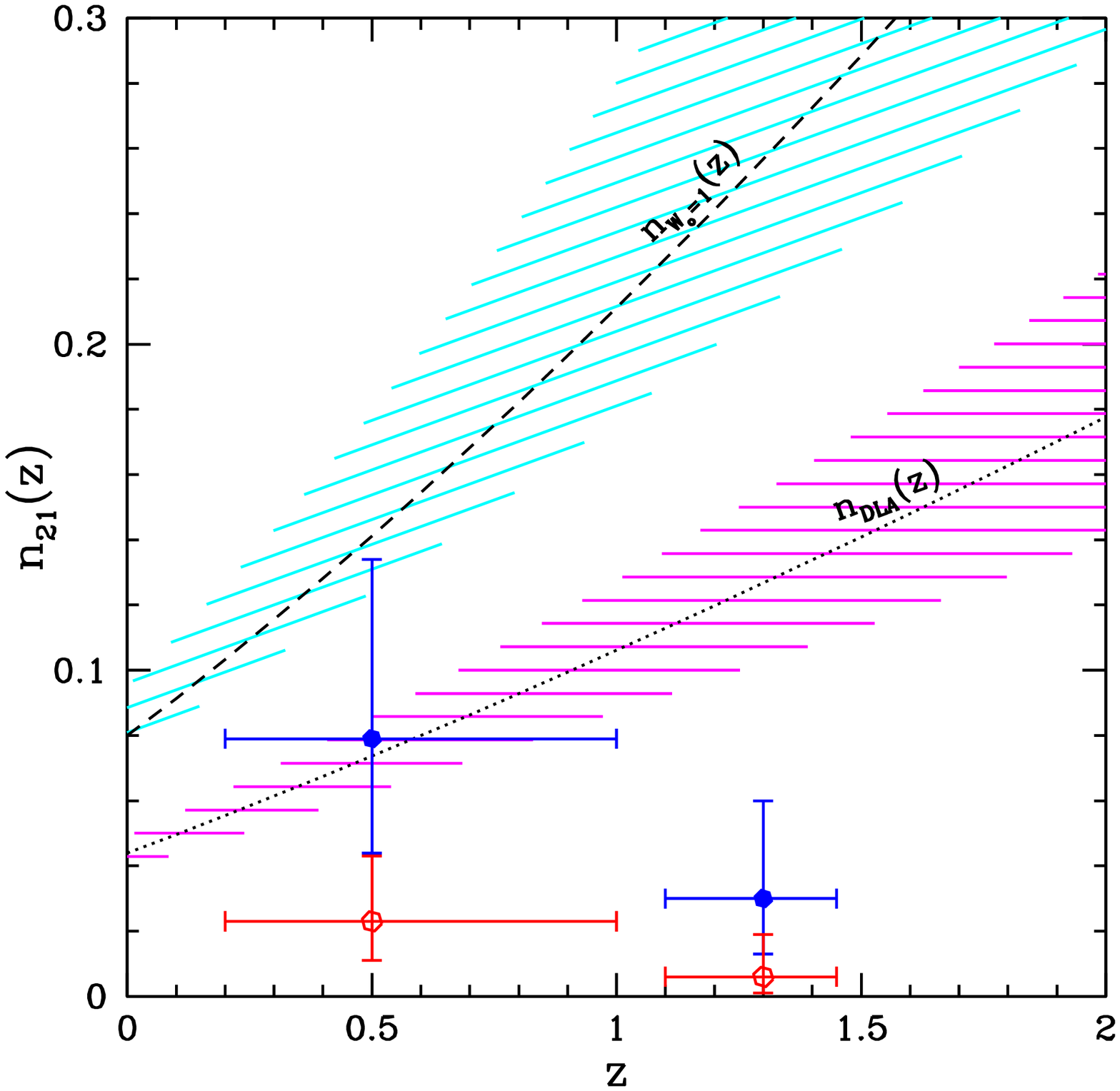,height=9.0cm,width=9.0cm,angle=0}
\caption{Number of 21-cm absorbers per unit redshift, $n_{21}(z)$, for 
integrated 21-cm optical depth, ${\cal{T}}_{\rm o}>0.3$ and 
$W_{\rm o}$(Mg~{\sc ii})~$>$~\,1.0\AA~ (solid symbols) and $W_{\rm o}$(Mg~{\sc ii})~$>$~1.8\,\AA~ (open symbols). 
Lines and hashed areas show the number of absorbers per unit redshift for DLAs (RTN06, dotted line) and
Mg~{\sc ii} absorbers with $W_{\rm o}\ge$1\AA~ (P06, dashed line).
}
\label{dndz}
\end{figure}



The number density per unit redshift $n_{21}({\cal{T}}_{\rm 0},W_{\rm o},z)$, of 21-cm systems with integrated
optical depth ${\cal{T}}_{21}\ge {\cal{T}}_{\rm 0}$ and Mg~{\sc ii} equivalent width
$W_{\rm r}\ge W_{\rm o}$ 
can be obtained from the number density per unit redshift of Mg~{\sc ii} absorbers, $n_{\rm Mg{\sc II}}(W_{\rm o},z)$, 
using the equation,
\begin{equation}
n_{21}({\cal{T}}_{21}\ge {\cal{T}}_0, W_{\rm r}\ge W_{\rm o}, z)= C \times n_{\rm Mg{\sc II}}(W_r\ge W_o, z).
\end{equation}
Where, $C$ is the fraction of Mg~{\sc ii} systems with $W_{\rm r}$(\mgiia)$\ge$$W_{\rm o}$ 
that show detectable 21-cm absorption. 
%

The redshift incidence of \mgii\, absorbers can be approximated as, 
\begin{equation}
n_{\rm Mg{\sc II}}(W_{\rm o},z) = n_0 \times (1+z)^\gamma
\end{equation}
with $n_0 = 0.080^{+0.015}_{-0.005}$ and $\gamma = 1.40\pm0.16$ for $W_{\rm o}=1$\AA~
and $n_0 = 0.016^{+0.005}_{-0.003}$ and $\gamma = 1.92\pm0.30$ for $W_r\ge 1.8$\AA~
(Prochter et al. 2006).
Therefore at $<$\zabs$>$=1.3, the mean redshift of our sample, 
we derive, $n_{\rm Mg{\sc II}}(W_{\rm o}= 1\AA,z=1.3)$ = $0.257^{+0.092}_{-0.046}$ and
$n_{\rm Mg{\sc II}}(W_{\rm o}= 1.8\AA,z=1.3)$ = $0.079^{+0.054}_{-0.030}$.

The fraction of \mgii systems that show detectable 21-cm absorption with integrated
optical depth, ${\cal{T}}_{21} \ge {\cal{T}}_0$, 
is estimated as follows. First we consider only the systems (say N) with radio spectra 
sensitive enough so that 
$\int\tau_{3\sigma}$dv$\le {\cal{T}}_0$.
Since the typical resolution of the Lane's spectra is 7-10\,km\,s$^{-1}$ (see below), we have computed the 
values of the 21-cm 3$\sigma$-detection limit 
for a velocity resolution of 10\,km\,s$^{-1}$ (see Table~\ref{mg2obsres}).  
Then the number of these systems with detectable 21-cm absorption  
(referred to as N$_{21}$) is estimated by considering the 
detections that have integrated line depth, ${\cal{T}}_{21} \ge {\cal{T}}_0$.  
Then $C$ is given by N$_{21}$/N and the error in $C$ is computed
using small number Poisson statistics (Gehrels, 1986) . 
The value of $n_{21}$ for various sub-samples defined using ${\cal{T}}_0$ and $W_{\rm o}$ are
summarised in Table~\ref{tabdndz}.
For $W_{\rm o}$ = 1 {\AA} the sub-sample with ${\cal{T}}_0 = 0.5$ contains 33 \mgii systems.
About 12\% of these systems are 21-cm absorbers. If we restrict ourselves
to compact QSOs with VLBA measurements, and use ${\cal{T}}_0$ corrected
for partial coverage (see Table.~\ref{vlbafc}),  we find that 17\% of 
these systems are 21-cm absorbers. Thus uncertainties due to unknown covering factor 
should have little effect on our conclusions.

%
We use the sample of Lane (2000, see Table~\ref{lanesamp})
to study the redshift evolution of $n_{21}$.
For two of the systems in this sample (\zabs = 0.524 towards 0827+243 and \zabs = 0.5313 towards 1629+120)
we use the results of better quality GMRT observations by Kanekar et al. (2003).
In the case of the system at \zabs = 0.6560 towards 1622+238, the 21-cm absorber detected by
Curran et al. (2007) is most probably located in front of the extended lobes.
As we have avoided such systems in our sample, we do not include it in our statistical analysis.
The values of $n_{21}$ for this sub-sample with $<z>\sim$0.5 are given in Table~\ref{tabdndz}. 

From Table 4 of Prochter et al. (2006) we expect the number density of Mg~{\sc ii} systems
with $W_{\rm o}$ = 1.0 and 1.8~\AA~ to increase by a factor of 1.8 and 2.3
respectively between $z=0.5$ and 1.3. If physical conditions
in Mg~{\sc ii} absorbers (like $N$(H~{\sc i}) and CNM fraction) do not
change with redshift we expect $C$ to remain constant and therefore
$n_{21}$ to increase with redshift. This is not the case
however. It is clear from Table~\ref{tabdndz} that, for a
given value of $W_{\rm o}$ and ${\cal{T}}_{\rm o}$, $C$ is less at higher $z$.
This decrease is even more pronounced for higher equivalent width 
($W_{\rm r}$(\mgiia)$\ge$1.8\AA). As the decrease in $C$ is
stronger than the increase in the number of Mg~{\sc ii} systems
per unit redshift interval we find $n_{21}$ decreases with redshift.

Rao, Turnshek \& Nestor (2006) give the number of DLAs : 
$n_{\rm DLA} (z=0.609) = 0.079\pm0.019$  and   $n_{\rm DLA}(z=1.219) = 0.120\pm0.025$ suggesting a
factor 1.5 increase in the number per unit redshift of DLAs. The number of 21-cm absorbers
with $W_{\rm o} > 1$\AA~ and ${\cal{T}}_{\rm o} > 0.3$ matches well the number of DLAs at $z\sim0.5$ 
but it is a factor of 4 smaller at $z \sim 1.3$ (see Fig.~\ref{dndz}).

The decrease in $n_{21}$ with redshift can be a consequence of the 
evolution of H~{\sc i} column density in the Mg~{\sc ii} absorbers or lower
CNM covering factor at high redshift which may reflect a strong evolution
either in the filling factor of the neutral cold gas or the $f_c$
of the radio sources.
From Table~\ref{tabdndz}, it is apparent that the fraction of 21-cm absorbers  
is not larger for $W_{\rm r}$(Mg~{\sc ii}$\lambda$2796)$\ge$1.8 \AA~ compared to
$W_{\rm r}$(Mg~{\sc ii}$\lambda$2796)$\ge$1 \AA~ at $z\sim 1.3$.
This is counter-intuitive as RTN06 have found that the fraction of DLAs amongst Mg~{\sc ii}
systems increases with increasing $W_{\rm r}$(Mg~{\sc ii}) (see their Fig.~6). 
Now the ${\cal{T}}_{\rm o}$~=~0.3 corresponds to log~$N$(H~{\sc i})~=~19.74 for the 
$T_{\rm s}$ = 100 K.  From the sample of RTN06 we find that the fraction of W$_{\rm r}\ge1\AA\,$ 
systems with log~$N$(H~{\sc i})~$\ge$~19.74 for the \zabs$\le$1.0 and \zabs$>$1.0 
is 50$\pm$10\% and 75$\pm$15\% respectively.
Thus the decrease in n$_{21}$ for ${\cal{T}}_{\rm o}$~=~0.3 is unlikely to be due to 
the evolution in column density amongst the Mg~{\sc ii} absorbers.
Further we have shown that the fraction of 21-cm absorbers in our
sample does not depend strongly on radio morphology (see Table~\ref{tabdndz}).
The available data are therefore consistent with the CNM fraction of neutral H~{\sc i} in Mg~{\sc ii}
systems being smaller at higher redshifts.
This will also be consistent with the results of M\'enard et al. (2008) who found the 
dust content of the strong Mg~{\sc ii} absorbers to be increasing with the 
cosmic time over the redshift range of $0.4 < z < 2.2$.
%

\section{Velocity spread of 21-cm absorption}

\begin{figure}
\psfig{figure=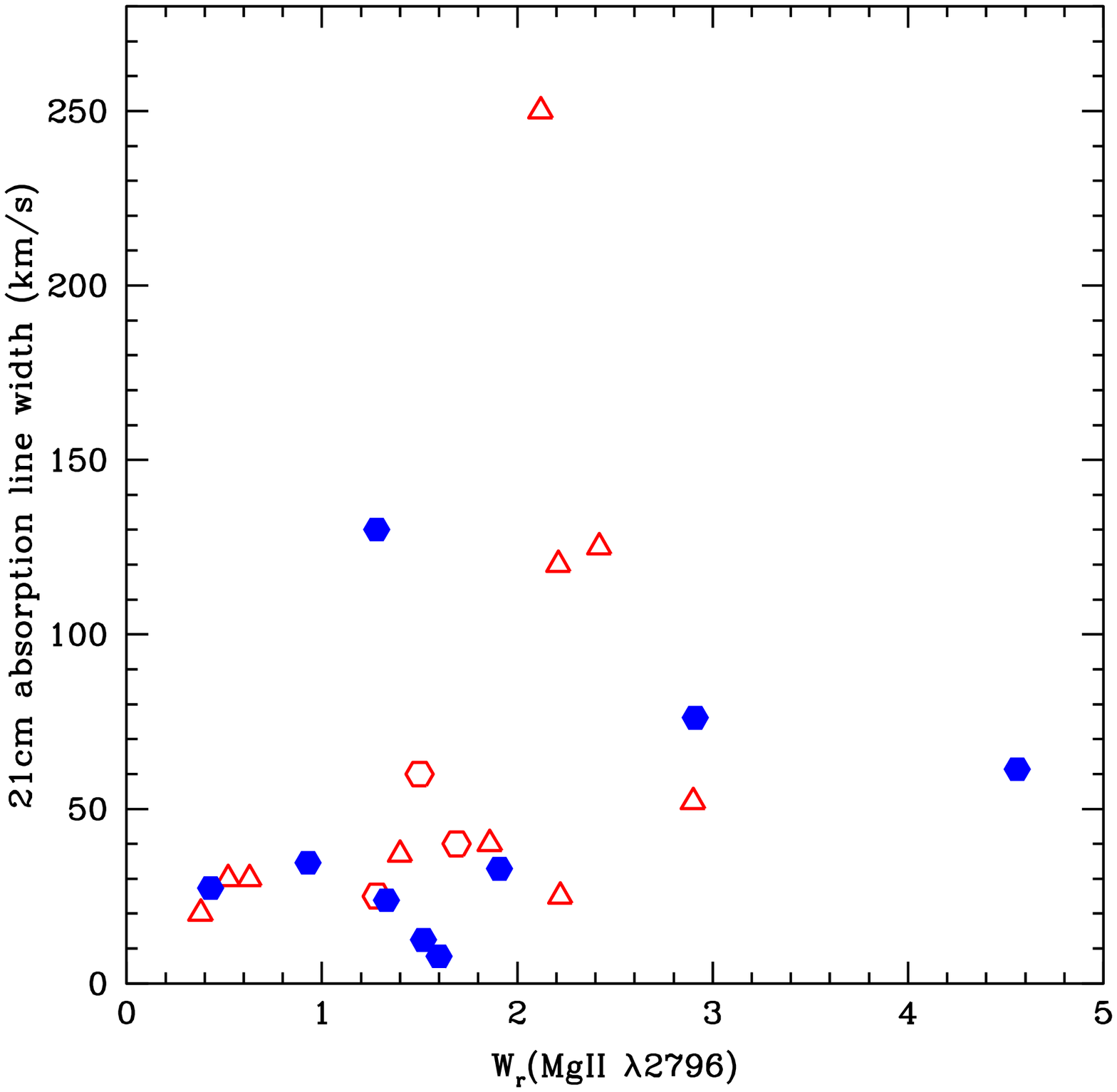,height=9.0cm,width=9.0cm,angle=0}
\caption{Width of the 21-cm absorption as a function of the
Mg~{\sc ii}$\lambda$2796 rest equivalent width.
Filled hexagons are for our GMRT sample, open triangles for the low-$z$
sample and open hexagons for  $z>1.8$ DLAs.  
}
\label{aopvw}
\end{figure}


In this Section we investigate any possible correlation between the 
velocity width of the 21-cm absorption and the Mg~{\sc ii}$\lambda2796$ rest equivalent width.
Indeed, Lane (2000) suggested that systems with large metal line equivalent widths could be
associated with multi-component 21-cm absorption. 
In addition, using mostly the sample described in Lane (2000),
Curran et al. (2007) have reported 
a correlation (significant at the  2.2$-$2.8$\sigma$ level) between 
the 21-cm line width and the Mg~{\sc ii}$\lambda$2796 equivalent width. 
%

\begin{table}
\caption{Results of non-parametric and rank correlation tests.}
\begin{tabular}{cccccc}
\hline
\hline
Sample$^*$         & $\rho$ & $\tau$ & \multicolumn{2}{c}{Probability } &$\sigma$\\
               &        &        & $\rho$ & $\tau$ & \\
(1)  & (2) & (3) & (4) & (5) \\
\hline
GMRT(9)           &0.133   & 0.056  & 0.732   & 0.835& 0.4  \\
GMRT-J0804+3012(8)&0.429   & 0.214  & 0.289   & 0.457& 1.0  \\
GMRT +high-$z$(12)&0.165   & 0.092  & 0.608   & 0.676& 0.6  \\
low-$z$(10)       &0.596   & 0.494  & 0.069   & 0.046& 1.8  \\
low-$z$ -0809+483 &0.678   & 0.592  & 0.045   & 0.026& 2.0  \\
All(22)           &0.482   & 0.337  & 0.023   & 0.028& 2.3  \\ 
All-0804-0809(20) &0.573   & 0.411  & 0.008   & 0.011& 2.8  \\
\hline
\end{tabular}
\begin{flushleft}
$^*$Numbers in brackets correspond to the number of 
21-cm absorbers in the subsamples.
\end{flushleft}
\label{tabcor}
\end{table}

In order to probe the presence of such a correlation in our sample 
we computed the 21-cm velocity width using the apparent optical 
depth profile and the method  described in Ledoux et al. (2006) 
for the low ionization metal lines (see Section~5 and Fig.~\ref{aop}). 
Even though our parent sample of Mg~{\sc ii} systems has a wide range in 
$W_{\rm r}$(Mg~{\sc ii}$\lambda$2796) most of our 21-cm detections correspond to 
systems with widths of 1-2 \AA. However, we find that measured 21-cm absorption 
velocity width ranges from 7 to 130 km/s (Table 4).

In Fig.~\ref{aopvw} we plot the 21-cm line width as a function of 
$W_{\rm r}$(Mg~{\sc ii}$\lambda$2796). Filled hexagons are for our GMRT sample, 
open triangles for the low-$z$ sample and open hexagons for $z>1.8$ DLAs. 
Results of non-parametric and rank correlation tests are 
summarized in Table~\ref{tabcor}. $\rho$ and $\tau$ are
the Spearman and Kendall rank correlation coefficients 
respectively. The probability for the measured correlation to
happen by chance is given in the 4th and 5th columns.  
The last column gives the significance of the correlation assuming the statistics 
to be Gaussian for the spearman rank correlation test.

We do not find any significant correlation 
between $W_{\rm r}$ and the 21-cm velocity width in the GMRT sample.
The largest 21-cm velocity spread is seen towards J0804+3012. Since the velocity spread
could be due to the complex structure of the radio source
that is resolved in our GMRT map (see discussions in Section 5) we 
removed this system from the analysis. The possible correlation is stronger
but still not significant ($\lapp 1\sigma$ assuming statistics to be Gaussian). 
Adding 3 high-$z$ 21-cm absorbers (open hexagons in
Fig.~\ref{aopvw}) observed at $1.8<z<2.1$ for which Mg~{\sc ii} equivalent widths are 
known does not modify the original conclusion of no correlation ($\lapp 1\sigma$).  
These 3 21-cm absorbers are \zabs=2.0394 towards 0458$-$020 (Wolfe et al. 1985; Wolfe et al. 1993), 
\zabs=1.9440 towards 1157+014 (Wolfe \& Davis 1979) and \zabs=1.7763 towards 1331+170 
(Wolfe \& Davis 1979; Briggs et al. 1983). 
%

The results for the low-$z$ data are plotted as open triangles. 
Here, we use the 6 21-cm absorbers from the low-$z$ sample 
discussed in the previous Section (see Table.~\ref{lanesamp})
plus 4 additional 21-cm absorption systems with Mg~{\sc ii} equivalent 
width measurements.
These four additional systems are \zabs=0.2213 towards 0738+313 (Lane et al. 1998; 
Kanekar et al. 2001), 
\zabs=0.4368 towards 0809+483 (Aldcroft et al. 1994; Briggs et al. 2001), 
\zabs=0.2377 towards 0952+179 (Lane 2000; Kanekar et al. 2001) and 
\zabs=0.692 towards 1328+307 (Davis \& May 1978; Cohen et al. 1994).
The correlation in this low-$z$ sample (with or without 0809+483 which has a 21-cm velocity width 
of 150\,km\,s$^{-1}$) is $\sim$2$\sigma$ significant.
There is therefore a difference between low and high redshift. 
The weak correlation (2$-$2.5 $\sigma$ level for a Gaussian
distribution) noted by Curran et al (2007) is 
present when we consider the full-sample i.e. all the sources 
at low and high redshift together.

The main differences between low-$z$ and GMRT samples
are (i) a paucity of 21-cm absorption systems with 
$W_{\rm r}$(Mg~{\sc ii}$\lambda$2796)$<$1 {\AA} 
and (ii) a lack of 21-cm absorbers among Mg~{\sc ii} systems with 
$W_{\rm r}$(Mg~{\sc ii})$>$1.8 {\AA} in the GMRT sample. While the first
difference is probably related to our sample selection the second difference
is probably due to a true evolution in the physical state of the
Mg~{\sc ii} systems with redshift (see discussion in the previous Section). 
Therefore we think it is not wise to mix  the absorbers at low and high redshifts.  

The absence of strong correlation between the total Mg~{\sc ii} equivalent 
width and the 21-cm absorption width is consistent with the idea that
the Mg~{\sc ii} equivalent width is mostly correlated with the overall kinematics of the
gas in the absorbing system and not with the column density in the
component associated with the cold gas.
In the case of J2340-0053 where we have access to high resolution 
echelle spectra, we find that the 21-cm absorption is not associated with
the strongest metal line component (see Fig. 9). In addition
as pointed out by Gupta et al. (2007) we usually find that the 21-cm
absorption redshift is slightly different (within 70 km~s$^{-1}$) from
the mean redshift defined by the metal absorption lines in the
SDSS spectrum. This can again be explained if the 21-cm absorption
does not originate from the strongest metal line components that
define the absorption redshift at low resolution.

\section{Summary and discussion}

We have reported the results of a systematic GMRT survey of 
21-cm absorption in a representative sample of strong Mg~{\sc ii} systems 
selected from the SDSS spectra in the redshift range 1.10$\le z \le$1.45. 
We have performed GMRT observations of $\sim$70\% (resp. 35\%) of our initial sample 
of Mg~{\sc ii} systems detected in DR5 with $W_{\rm r}$(Mg~{\sc ii}$\lambda$2796)$\ge$1 {\AA} 
located in front of compact radio sources with flux density $\ge$100\,mJy (resp. in the range 50-100\,mJy).
We show that our observed sample is representative of the full parent population.

We report detection of 9 new 21-cm absorption systems out of 35 systems observed in our 
survey. This is by far the largest number of systems detected in a single systematic 
survey in a narrow redshift range. Two of these systems also show 2175~{\AA} dust feature
at the redshift of the absorbers (Srianand et al. 2008b).
We provide spectra of all non-detections with upper limits on the 21-cm optical depth. 

We study the dependence of detectability of 21-cm absorption on different 
properties of the UV absorption lines detected in the SDSS spectra.
We find that if absorption systems are selected with a Mg~{\sc ii} doublet ratio, DR~$<$~1.3, and 
a ratio $W_{\rm r}$(Mg~{\sc i})/$W_{\rm r}$(Mg~{\sc ii})~$>$~0.3, the success rate 
for 21-cm detection is very high (up to $90$\%).
We notice that the detections found in a low-$z$ sample by Lane (2000) also obey 
these joint constraints. Low value of DR and 
high values of $W_{\rm r}$(Mg~{\sc i})/$W_{\rm r}$(Mg~{\sc ii}) in a system
will mean a high column density of Mg~{\sc ii} and high neutral fraction. 
Thus this selection favors objects with high N(H~{\sc i}).
In our sample, we find an apparent paucity of 21-cm absorption among
systems with $W_{\rm r}$(Mg~{\sc ii}$\lambda$2796)$>$1.8 {\AA}, 
the median $W_{\rm r}$ of our sample. This is contrary to what has been seen at
low-$z$ (Lane 2000).
Interestingly most of these high $W_{\rm r}$ systems 
have high DR and low values of $W_{\rm r}$(Mg~{\sc i})/$W_{\rm r}$(Mg~{\sc ii}).
This strongly suggests that the equivalent width in these systems is dominated
by velocity spread and not by line saturation. 

We discuss the number of 21-cm absorption systems per unit redshift interval 
for a given limiting value of the integrated 21-cm optical depth
and $W_{\rm r}$(Mg~{\sc ii}$\lambda$2796). 
We show that  
the fraction of Mg~{\sc ii} systems with 21-cm absorption and the
$n_{21}$ decrease from $z\sim0.5$ to $z\sim1.3$.
The decrease is larger when we use higher equivalent width cutoff. 
Using a sub-sample of compact sources, with high frequency VLBA observations 
available, we show that this can not be accounted for by simple covering factor effects. 
As mentioned above and based on the available data, it appears that most likely 
the main reason behind this cosmological evolution is the decrease of the CNM covering factor
(and volume filling factor) in the strong Mg~{\sc ii} absorbers.
Indeed, it is known that the number of Mg~{\sc ii} systems per unit redshift increases 
with increasing redshift. The evolution is steeper for stronger systems
(Steidel \& Sargent, 1992 and P06 for recent reference).
Using the data of Steidel \& Sargent (1992), Srianand (1996) 
found that the strongest redshift evolution was seen among the 
Mg~{\sc ii} absorbers with $W_{\rm r}$(Fe~{\sc ii})$\lambda$2383/$W_{\rm r}$(Mg~{\sc ii})$\lambda$2796$<$0.5. 
This clearly means the physical conditions in strong Mg~{\sc ii} absorbers are
different at high and low-$z$.

Previous surveys (listed in Section 2) have hinted the lack of 21-cm
absorption systems at high-$z$ ($z\ge 2$) compared to what is seen at low-$z$, $z\le$0.5.  
The low detection rate of 21-cm absorption in high-$z$ DLAs is attributed 
to either the gas being warm (high $T_{\rm s}$) or to low values of 
the covering factor because of high-$z$ geometric effects (see Kanekar \& Chengalur, 2003
and Curran \& Webb, 2006, respectively). In two of our systems 
(at \zabs = 1.326 towards J0850+5159 and \zabs = 1.209 towards J0852+3435), the presence of 
a 2175{\AA} dust feature allows us to derive an estimate of $N$(H~{\sc i})
and therefore $T_{\rm s}$ (Srianand et al. 2008b). 
Both systems have high metallicity, 
large $N$(H~{\sc i}) ($\sim$6$\times$10$^{21}$~cm$^{-2}$) and  $T_{\rm s}$
consistent with the temperature in cold neutral (CNM) gas. In the case of the system at \zabs = 1.361 
towards J2340$-$0053, the 21-cm line 
width is consistent with the absorbing gas having $T\le200$~K. Therefore, we already have 
3 systems in our sample with properties consistent with the conditions expected for
CNM gas. Only 1 such system has yet been detected at $z\ge 2$ (York et al. 2007). 
Thus our results are consistent with the increase of the CNM covering factor with decreasing
redshift.
However, to ascertain this result 
it would be important to derive $N$(H~{\sc i}) directly from the Lyman-$\alpha$ absorption. 
Thus followup HST/COS observations of a UV bright subsample of our sample is
very important to study the redshift evolution of $T_{\rm s}$.
%
As pointed out by numerous authors, the physical state of the gas depends on
the star formation activity in the vicinity of the absorber. 
Models of multiphase interstellar medium suggest that the temperature of 
different phases depends on the background UV radiation field, metallicity, dust content and 
cosmic ray density and therefore on the local star formation activity.
It is possible to conjecture that the evolution of $n_{21}$ as a function of redshift 
is probably related to the fact that at $z\sim1.3$ the average SFR in a galaxy is larger than 
that at $z\sim0.5$ so that we expect the CNM fraction to increase with
decreasing redshift. 
%

%
We selected systems with $W_{\rm r} \ge$1\,\AA~ but detected by chance a 21-cm absorption
in a system with  $W_{\rm r}$~=~0.43\,\AA (at $z_{\rm abs}$~=~1.3710 toward J0108$-$0037).
Efforts are underway at GMRT to extend our survey to weaker ($W_{\rm r} \le$1\,\AA~) Mg~{\sc ii} systems. 
This will be crucial for understanding the physical state of Mg~{\sc ii} systems and 
to determine the detectability of 21-cm absorption versus $W_{\rm r}$.  
Ideally one would like to estimate the number density of 21-cm absorbers and measure the cosmological 
evolution without preselection from the UV absorption lines. This can be achieved only by a blind 
survey of 21-cm absorption in front of radio loud QSOs. 
It will be possible to embark upon 
such a survey with the upcoming Square Kilometer Array (SKA) pathfinders.
In particular, the Australian Square Kilometer Array Pathfinder (ASKAP) with its  instantaneous wide bandwidth 
of 300\,MHz and large field of view (30\,degree$^2$) is an ideal instrument for this
(Johnston et al. 2008). 
An ASKAP survey with 150 pointings of 16~hrs each (i.e 2400 hrs in total) in the 700-1000 MHz frequency 
band would yield detection of $\sim$100 to 250 intervening 21-cm absorbers in the redshift range 0.4$\le z \le$1. 
%

We have estimated the velocity spread of the 21-cm absorptions using the 
apparent optical depth method (Ledoux et al. 2006). 
We do not find any statistically significant correlation between 
$W_{\rm r}$(Mg~{\sc ii}$\lambda$2796) and the 21-cm velocity width in our sample. 
A marginal correlation is found for the low-$z$ sample.
The absence of correlation in the high-$z$ sample is related to
the lack of 21-cm absorbers among Mg~{\sc ii} systems with 
$W_{\rm r}$(Mg~{\sc ii})$>$1.8 {\AA} in the GMRT sample. This
is probably due to a true evolution with redshift of the physical state of the
Mg~{\sc ii} systems and consistent with the idea that
the Mg~{\sc ii} equivalent width is mostly correlated with the overall kinematics of the
gas in the absorbing system and not with the column density in the
component associated with the cold gas.
When high spectral resolution data are available, we note that the 21-cm absorption 
is not always associated with the strongest Mg~{\sc ii} component.

As the energy of the 21-cm transition is proportional to $x=\alpha^2 G_p/ \mu$, 
high resolution optical and 21-cm spectra can be used together to probe the combined
cosmological variation of these constants (Tubbs \& Wolfe, 1980).
Using this technique on scanned data, Tzanavaris et al. (2005) obtained
${\Delta x/ x} = (0.63\pm0.99)\times 10^{-5}$.
Our GMRT survey provides systems in a narrow redshift range in which this
measurement can be done.  
Thus high resolution optical spectroscopy of the corresponding QSOs are suitable
to perform this test at $z\sim1.3$.

\section{acknowledgements}
We thank Rajaram Nityananda for useful discussions and 
providing Director's discretionary time on several occassions. 
We thank the GMRT staff for their co-operation during our observations.
The GMRT is an international facility run by the National Centre for Radio 
Astrophysics of the Tata Institute of Fundamental Research.
We acknowledge the use of SDSS spectra from the archive (http://www.sdss.org/).
RS and PPJ gratefully acknowledge support from the Indo-French
Centre for the Promotion of Advanced Research.



\end{document}